\documentclass{ws-ijmpd}
\usepackage{cite}
\usepackage{bm}
\usepackage{slashed}
\usepackage{braket}
\usepackage{amsmath}
\usepackage{amsfonts}
\usepackage{mathrsfs}
\usepackage{url}

\def\draftnote{}
\def\trimmarks{}

\numberwithin{equation}{section}

\begin{document}
	
\markboth{Balakumar and Winstanley}
{Hadamard parametrix of the Green's function of a five-dimensional charged scalar field}

%
%

\title{Hadamard parametrix of the Feynman Green's function of a five-dimensional charged scalar field}

\author{VISAKAN BALAKUMAR}
\address{Consortium for Fundamental Physics, School of Mathematics and Statistics, \\ University of Sheffield,
	Hicks Building, Hounsfield Road, Sheffield. S3 7RH United Kingdom \\
VBalakumar1@sheffield.ac.uk}

\author{ELIZABETH WINSTANLEY}

\address{Consortium for Fundamental Physics, School of Mathematics and Statistics, \\ University of Sheffield,
	Hicks Building, Hounsfield Road, Sheffield. S3 7RH United Kingdom \\
E.Winstanley@sheffield.ac.uk}

\maketitle


\begin{abstract}
The Hadamard parametrix is a representation of the short-distance singularity structure of the Feynman Green's function for a quantum field on a curved space-time background. 
Subtracting these divergent terms regularizes the Feynman Green's function and enables the computation of renormalized expectation values of observables. 
We study the Hadamard parametrix for a charged, massive, complex scalar field in five space-time dimensions.
Even in Minkowski space-time, it is not possible to write the Feynman Green's function for a charged scalar field exactly in closed form.
We therefore present covariant Taylor series expansions for the biscalars arising in the Hadamard parametrix. On a general space-time background, we explicitly state the expansion coefficients up to the order required for the computation of the renormalized scalar field current.
These coefficients become increasingly lengthy as the order of the expansion increases, so we give the higher-order terms required for the calculation of the renormalized stress-energy tensor in Minkowski space-time only.
\end{abstract}




\section{Introduction}
\label{sec:intro}

Finding a definitive theory of quantum gravity, in which the gravitational field and matter are fully quantized, remains one of the most important open questions in fundamental physics (see, for example, \cite{Kiefer:2005uk} for a review).
One avenue to elucidating some of the features of a full theory of quantum gravity is to take a semiclassical approach, namely quantum field theory on curved space-time.
In this set-up, the gravitational field and space-time geometry remain fixed and classical, with only the matter fields quantized (see, for example, \cite{DeWitt:1975ys,Birrell:1982ix,Fulling:1989nb,Wald:1995,Parker:2009uva} for reviews). 
This approach is informative because a theory of quantum gravity, if it is to be successful, must give results identical to those of quantum field theory on curved space-time in an appropriate limit. 
Quantum field theory on curved space-time has also revealed many deep and significant physical effects, including the creation of particles in an expanding universe  \cite{Parker:1968mv,Parker:1969au,Parker:1971pt}, the Unruh effect \cite{Fulling:1972md,Davies:1974th,Unruh:1976db,Crispino:2007eb} and the Hawking radiation of black holes \cite{Hawking:1974rv,Hawking:1974sw}. 

In order to study the physics of a quantum field theory on a particular curved space-time, it is useful to compute expectation values of observables in a chosen quantum state.
For example, the stress-energy tensor $T_{\mu \nu }$ contains valuable information about the energy density and fluxes of the quantum field.
Expectation values of observables typically involve products of field operators at the same space-time point.
Even for the simplest type of quantum field, a free neutral scalar field, on the simplest space-time, namely Minkowski space-time, such expectation values are formally infinite.
For a free quantum field on flat space-time, this is easily rectified by setting the expectation value of $T_{\mu \nu }$ to vanish in the global Minkowski vacuum, and effectively considering differences in expectation values between a chosen quantum state and the vacuum.
As we shall explain later, for a quantum field on curved space-time, these infinities are not so straightforwardly regularized.

Many different approaches to dealing with these infinities in expectation values have been developed, including De-Witt Schwinger, dimensional, Pauli-Villars, adiabatic and zeta-function regularization  
\cite{DeWitt:1975ys,Birrell:1982ix,Fulling:1989nb,Wald:1995,Parker:2009uva}.
Expectation values of observables such as the stress-energy tensor can be computed from the Feynman Green's function of the quantum field, so the problem of regularizing expectation values can be solved by regularizing the Feynman Green's function.
In this paper, we focus on one representation of the short-distance singularity structure of the Feynman Green's function, namely the Hadamard parametrix.
Physically reasonable quantum states have a Feynman Green's function with this singularity structure \cite{Fewster:2013lqa}.
Finding the Hadamard parametrix for a quantum field on a curved space-time background is therefore the first step to regularizing the Feynman Green's function and  hence computing expectation values of observables.

In this report we study the Hadamard parametrix of the Feynman Green's function for a charged, massive, complex scalar field with arbitrary coupling to the space-time curvature.
We begin, in section \ref{sec:charged}, by briefly outlining the key equations of such a charged scalar field theory in $d$ space-time dimensions, including expressions for the expectation values of observables in terms of the Feynman Green's function.
For the rest of the paper, we restrict our attention to $d=5$ space-time dimensions.
This case is of particular interest for Kaluza-Klein theory \cite{Kaluza:1984ws,Klein:1926tv}, as well as brane-world \cite{ArkaniHamed:1998rs,Antoniadis:1998ig,ArkaniHamed:1998nn}
and  Randall-Sundrum scenarios \cite{Randall:1999ee,Randall:1999vf}.
The Hadamard parametrix of the Feynman Green's function in five dimensions is outlined in section \ref{sec:Hadamard}, following our previous work \cite{Balakumar:2019djw}. 
The Hadamard parametrix depends on a set of biscalars which cannot be determined in closed form.
In section \ref{sec:Ucalc} covariant Taylor series expansions of these biscalars are derived in detail.
We work to the order required for the computation of the renormalized stress-energy tensor, but present explicit general expressions for the expansion coefficients only up to the order required for the computation of the renormalized current.
The higher-order terms needed for finding the renormalized stress-energy tensor are extremely lengthy, so we give them only on Minkowski space-time.
These results extend the work of \cite{Balakumar:2019djw}, where the covariant Taylor series expansions were given for $d=2,3,4$.
Our conclusions are presented in section \ref{sec:conc}.

\section{Charged scalar field theory}
\label{sec:charged}

In $d$-dimensional flat space-time, a neutral scalar field $\Phi $ of mass $m$ is governed by the Klein-Gordon equation
\begin{equation}
   \left[ \partial _{\mu }\partial ^{\mu } - m^{2}  \right] \Phi =0.
   \label{eq:KGflatneutral}
\end{equation}
In this model, the scalar field $\Phi $ takes real values and we are considering only a free scalar field with no self-interaction potential. 
It is also possible to consider a free complex scalar field $\Phi $ satisfying (\ref{eq:KGflatneutral}). 
The model can be extended to a complex scalar field with charge $q$, minimally coupled to an electromagnetic field  $F_{\mu \nu }$ derived from an electromagnetic potential $A_{\mu }$ by replacing the partial derivatives $\partial _{\mu } $ with gauge covariant derivatives $\partial _{\mu } - {\rm {i}}q A_{\mu }$.
In this paper we make a further generalization, and  consider a charged complex scalar field on a general $d$-dimensional curved space-time background, in which case the partial derivatives $\partial _{\mu }$ are replaced by space-time covariant derivatives $\nabla _{\mu }$ and (\ref{eq:KGflatneutral}) becomes
\begin{equation}
    \left[ D_{\mu }D^{\mu } - m^{2} -\xi R \right] \Phi =0,
    \label{eq:KGcharged}
\end{equation}
where $D_{\mu }= \nabla _{\mu } - {\rm {i}}qA_{\mu }$.
In (\ref{eq:KGcharged}) we have included a nonmiminal coupling between the scalar field and the space-time curvature scalar $R$, where $\xi $ is the coupling constant. If we set $\xi =0$, the scalar field is minimally coupled to the space-time curvature.
Here and throughout this paper we use units in which $8\pi G = c = \hbar =1$ and the space-time metric has mostly plus signature. 

We now consider the situation in which the scalar field ${\hat {\Phi  }}$ has been quantized but the background space-time geometry, given by the metric $g_{\mu \nu }$, and the electromagnetic potential $A_{\mu }$ are fixed and classical. 
Thus our model is a version of scalar QED, as has been studied recently on cosmological space-times in two \cite{Ferreiro:2018qzr} and four \cite{Ferreiro:2018qdi} dimensions.
In any quantum field theory, one is interested in the computation of expectation values of observables.
In our model, observables of interest are the scalar field condensate, the current ${\hat {J}}^{\mu }$ and the stress-energy tensor ${\hat {T}}_{\mu \nu }$. 

Given a particular quantum state, the expectation values of these observables can be computed from the Feynman Green's function for the charged scalar field in that state.
A Green's function for the charged scalar field equation (\ref{eq:KGcharged}) is a function $G(x,x')$, depending on two space-time points, which satisfies the inhomogeneous equation
\begin{equation}
   \left[ D_{\mu }D^{\mu } - m^{2}-\xi R \right] G(x,x') =
   - \left[ -g(x) \right] ^{-\frac{1}{2}} \delta ^{d} (x-x')
   \label{eq:inhom}
\end{equation}
where $g(x)$ is the determinant of the space-time metric and $\delta ^{d}(x-x')$ is the $d$-dimensional Dirac delta function.
There are various different Green's functions which satisfy (\ref{eq:inhom}), corresponding to different choices of contour in momentum space (see, for example, \cite{Birrell:1982ix} for details). 
In this article we consider the Feynman Green's function $G_{\rm {F}}(x,x')$, which corresponds to the following expectation value:
\begin{equation}
    -{\rm {i}}G_{\rm {F}}(x,x')= \left\langle T\left[ {\hat {\Phi }}(x) {\hat {\Phi }}^{\dagger }(x') \right] \right\rangle  .
    \label{eq:GFdef}
\end{equation}
Here $T\left[ {\hat {\Phi }}(x) {\hat {\Phi }}^{\dagger }(x') \right]$ denotes the time-ordered product 
\begin{equation}
T\left[ {\hat {\Phi }}(x) {\hat {\Phi }}^{\dagger }(x') \right] =
\Theta \left( x^{0}-x^{0'} \right) {\hat {\Phi }}(x) {\hat {\Phi }}^{\dagger }(x')
+
\Theta \left( x^{0'}-x^{0} \right) {\hat {\Phi }}^{\dagger } (x'){\hat {\Phi }}(x) ,
\label{eq:timeordered}
\end{equation}
where $\Theta (x)$ is the Heaviside step function, and ${\hat {\Phi }}^{\dagger }$ denotes the adjoint field operator (which is not equal to ${\hat {\Phi }}$ for a complex scalar field).
At first glance the definition (\ref{eq:timeordered}) depends on a choice of time coordinate.
However, if the field operators commute when the points $x$ and $x'$ are space-like separated, then the Feynman Green's function is a biscalar quantity, that is, a scalar function of both $x$ and $x'$.

The naive expectation value of the scalar field condensate corresponds to taking the limit $x'\rightarrow x$ in the Feynman Green's function:
\begin{equation}
\langle {\hat {\Phi }}{\hat {\Phi }}^{\dagger} \rangle _{\rm {ren}}
    = \lim _{x'\rightarrow x} \Re \left\{ -{\rm {i}}G_{\rm {F}}(x,x') \right\} ,
    \label{eq:condensate}    
\end{equation}
while the expectation values of the current and stress-energy tensor are, respectively,
\begin{equation}
    \langle {\hat {J}}^{\mu } \rangle =
    -\frac{q}{4\pi } \lim _{x'\rightarrow x} \Im \left\{ D^{\mu } \left[ - {\rm {i}}G_{\rm {F}}(x,x')  \right] \right\} ,
    \label{eq:currentdef}
    \end{equation}
    where $\Im $ denotes the imaginary part and
    \begin{equation}
    \langle {\hat {T}}_{\mu \nu}\rangle =
    \lim _{x'\rightarrow x}\Re\left\{  {\mathcal {T}}_{\mu \nu }(x,x') \left[ - {\rm {i}}G_{\rm {F}}(x,x') \right] \right\}  ,
    \label{eq:Tmunudef}
\end{equation}
where $\Re $ denotes the real part and ${\mathcal {T}}_{\mu \nu }$ is the second order differential operator
\begin{eqnarray}
{\mathcal {T}}_{\mu \nu } & = & 
\left( 1- 2\xi  \right) g_{\nu }{}^{\nu '} D_{\mu }D^{*}_{\nu '}
+ \left( 2\xi - \frac{1}{2} \right) g_{\mu \nu }g^{\rho \tau '}D_{\rho }D^{*}_{\tau '}
-2\xi  D_{\mu }D_{\nu }
\nonumber \\ & & 
+ 2\xi g_{\mu \nu }D_{\rho }D^{\rho } 
+ \xi \left( R_{\mu \nu }- \frac{1}{2}g_{\mu \nu }R \right) - \frac{1}{2}m^{2} g_{\mu \nu },
\label{eq:RSETop}
\end{eqnarray}
with $g_{\mu }{}^{\mu '}$ the bivector of parallel transport.
In (\ref{eq:RSETop}), we have $D^{*}_{\mu }=\nabla _{\mu }+{\rm {i}}qA_{\mu }$ and
the operator $D_{\nu '}$ acts at the space-time point $x'$.
The presence of the Dirac delta function on the right-hand-side of the governing equation (\ref{eq:inhom}) tells us that the Feynman Green's function is in fact singular in the limit $x'\rightarrow x$ and therefore the limits in 
(\ref{eq:condensate}--\ref{eq:Tmunudef}) do not yield finite quantities.
Therefore some method of regularization (identifying the singularities) and renormalization (removing these singularities to give finite expectation values) is required.

Both the Feynman Green's function $G_{\rm {F}}(x,x')$ and the expectation values (\ref{eq:condensate}--\ref{eq:Tmunudef}) are finite if the space-time points $x$ and $x'$ are separated.  
Therefore we follow the point-splitting approach to regularization \cite{DeWitt:1975ys,Christensen:1976vb,Christensen:1978yd}, considering $x$ and $x'$ to be closely separated, but distinct, and such that there is a unique geodesic connecting them.
Our main result in this paper is the derivation of an appropriate parametrix (the {\em {Hadamard parametrix}}) $G_{\rm {H}}(x,x')$ which has the same short-distance divergences as $G_{\rm {F}}(x,x')$.
For concreteness, in this article we focus on the case of a five-dimensional space-time (results for lower numbers of space-time dimensions can be found in \cite{Balakumar:2019djw}).
The next section outlines the form of $G_{\rm {H}}(x,x')$ for a charged scalar field in five space-time dimensions,
which depends on a set of biscalars, dubbed the {\em {Hadamard parameters}}. 
In section \ref{sec:Ucalc} we derive covariant series expansions for these Hadamard parameters. 
We shall see that $G_{\rm {H}}(x,x')$ is independent of the state of the quantum field, and depends only on the properties of the background space-time geometry and electromagnetic potential.

\section{Hadamard parametrix of the Feynman Green's function in five dimensions}
\label{sec:Hadamard}

In this section we study $G_{\rm {H}}(x,x')$ for a charged scalar field in five space-time dimensions.
We first consider the simpler system of a neutral scalar field on five-dimensional Minkowski space-time, for which $G_{\rm {H}}(x,x')$ can be given exactly in closed form.
This will enable us to examine the singularity structure in some detail and motivate the general form of $G_{\rm {H}}(x,x')$ for a charged scalar field on a curved space-time background.

\subsection{Neutral scalar field on five-dimensional flat space-time}
\label{sec:neutral5Dflat}

Consider first a neutral scalar field on five-dimensional Minkowski space-time, satisfying the Klein-Gordon equation (\ref{eq:KGflatneutral}).
The Feynman Green's function $G_{\rm {F}}^{\rm {M}}(x,x')$ then satisfies the inhomogeneous equation
\begin{equation}
   \left[ \partial _{\mu }\partial ^{\mu } - m^{2}  \right] G_{\rm {F}}^{\rm {M}}(x,x') =- \delta ^{5}(x-x').
   \label{eq:KGflatneutralinhom}
\end{equation}
Here and throughout this paper we use the label ${\rm {M}}$ to denote a quantity on Minkowski space-time.
In this case the Feynman Green's function can be found in closed form for all space-time points $x,x'$ \cite{Zhang2010}:
\begin{eqnarray}
    -{\rm {i}}G_{\rm {F}}^{\rm {M}}(x,x') 
    & = & 
-\frac{{\rm {i}}{\sqrt {m^{3}}}}{32{\sqrt {\pi ^{3}}}}
    \frac{1}{\left( \sigma _{\rm {M}}- {\rm {i}}\epsilon \right) ^{\frac{3}{4}} }
    H_{\frac{3}{2}}^{(2)}\left( m{\sqrt {\sigma _{\rm {M}}- {\rm {i}}\epsilon }}\right)
     +W_{\rm {M}}(x,x') ,
     \label{eq:GFflat}
\end{eqnarray}
where $\sigma _{\rm {M}}$ is one half the square of the distance between the two points in flat space-time
\begin{equation}
    \sigma _{\rm {M}} =\frac {1}{2} \eta _{\mu \nu } (x^{\mu }-x'^{\mu })(x^{\nu }-x'^{\nu }) ,
    \label{eq:sigmaflat}
\end{equation}
and $\eta _{\mu \nu }={\rm {diag}}\{ -1,1,1,1,1 \}$ is the five-dimensional Minkowski metric.
In (\ref{eq:GFflat}) we have assumed that $\sigma _{\rm {M}}>0$ and the points are space-like separated.
For simplicity of exposition, we shall assume space-like separation for the rest of this paper.
In addition, $H^{(2)}_{\frac {3}{2}}$ is a Hankel function of the second kind.
The quantity $\epsilon \rightarrow 0 $ is introduced in (\ref{eq:GFflat}) so that the Feynman Green's function has the correct analyticity properties and we shall set $\epsilon =0$ for the remainder of this paper.
Finally, $W_{\rm {M}}(x,x')$ is any solution of the homogeneous scalar field equation (\ref{eq:KGflatneutral}) and is therefore regular in the coincidence limit $x'\rightarrow x$.  
If the quantum scalar field is in the vacuum state, $W_{\rm {M}}$ vanishes identically \cite{Zhang2010}, but it is nonzero for a general quantum state.

The first term in (\ref{eq:GFflat}) is the same for all quantum states and is singular in the limit $x'\rightarrow x$. It is also singular when the points are null separated and $\sigma _{\rm {M}}=0$ (this singularity is regulated by the ${\rm {i}}\epsilon $ term), but our interest in this paper is in the short-distance singularity structure of the Feynman Green's function.
We therefore define the singular part of the Feynman Green's function to be
\begin{equation}
-{\rm {i}}G_{\rm {S}}(x,x')= -{\rm {i}}G_{\rm {F}}^{\rm {M}}(x,x')
- W_{\rm {M}}(x,x').
\end{equation}
The Feynman Green's function can be regularized by subtracting $G_{\rm {S}}(x,x')$ from $G_{\rm {F}}^{\rm {M}}(x,x')$, leaving the regular, state-dependent part equal to $W_{\rm {M}}(x,x')$.
Renormalized expectation values are then computed by replacing the Feynman Green's function by $W_{\rm {M}}(x,x')$.
This gives the renormalized expectation value of the scalar field condensate (\ref{eq:condensate}) to be $\lim _{x'\rightarrow x}\left[ W_{\rm {M}}(x,x')\right] $, and the renormalized expectation value of the stress-energy tensor is given by (\ref{eq:Tmunudef}) with the operator ${\mathcal {T}}_{\mu \nu }$ (\ref{eq:RSETop}) acting on $W_{\rm {M}}(x,x')$ rather than $-{\rm {i}}G_{\rm {F}}^{\rm {M}}(x,x')$.
Since for the moment we are considering only a neutral scalar field, the expectation value of the current (\ref{eq:currentdef}) vanishes identically for all quantum states.
When the quantum field is in the vacuum state and $W_{\rm {M}}(x,x')$ is zero, the expectation values of the scalar field condensate and stress-energy tensor are also zero, as expected.

For a general quantum state, the Feynman Green's function is typically given as a sum over mode solutions of the homogeneous scalar field equation (\ref{eq:KGflatneutral}). 
It can therefore be useful to apply the differential operators arising in (\ref{eq:currentdef}, \ref{eq:Tmunudef}) to the Feynman Green's function $G_{\rm {F}}^{\rm {M}}(x,x')$ before subtracting the divergent parts arising from the application of the operators to $G_{\rm {S}}(x,x')$.
To find these divergent parts, we do not need to consider the exact expression for $G_{\rm {S}}(x,x')$. 
Instead an expansion in $\sigma _{\rm {M}}$ to sufficiently high order will suffice.
The operator ${\mathcal {T}}_{\mu \nu }$ involves second order derivatives, so in order to find the divergent parts of the expectation value of the stress-energy tensor, we require the expansion of  $G_{\rm {S}}(x,x')$ to order ${\mathcal {O}}(\sigma _{M}^{2})$.
For a real scalar field, we only require the expansion of $\Re \left\{  -{\rm {i}} G_{\rm {S}} (x,x')\right\}$ since all expectation values are real.
For small $\sigma _{\rm {M}} >0$, this is given by
\begin{equation}
  \Re \left\{ -{\rm {i}} G_{\rm {S}} (x,x') \right\}  =  
    \frac{1}{16{\sqrt {2}}\pi ^{2}\sigma ^{\frac{3}{2}}_{\rm {M}}}
    \left[ 
    1 
    +\frac{m^{2}}{2}\sigma _{\rm {M}}
    -\frac{m^{4}}{8}\sigma _{\rm {M}}^{2} 
   +\frac{m^{6}}{144}\sigma _{\rm {M}}^{3}
    + {\mathcal {O}}(\sigma _{\rm {M}}^{4}) 
    \right] .
    \label{eq:Gcloseflat}
\end{equation}
The leading-order singularity in (\ref{eq:Gcloseflat}) is ${\mathcal {O}}(\sigma _{\rm {M}}^{-\frac{3}{2}})$ and we have obtained an expansion of the form $\sigma _{\rm {M}}^{-\frac{3}{2}}$ multiplied by a power series expansion in $\sigma _{\rm {M}}$. 
If the scalar field is massless, then the power series reduces to unity.

\subsection{Charged scalar field on five-dimensional curved space-time}
\label{sec:5dcurved}

We now return to our main model, namely a charged scalar field on a curved space-time background.
In this case, unlike the simpler flat space-time example considered in the previous subsection, it is not possible to derive the general solution of the inhomogeneous scalar field equation (\ref{eq:inhom}) in closed form.
However, the Feynman Green's function will still consist of two parts:  a particular solution $G_{\rm {H}}(x,x')$ of the inhomogeneous equation (\ref{eq:inhom}) together with the general solution $W(x,x')$ of the homogeneous equation (\ref{eq:KGcharged}):
\begin{equation}
    -{\rm {i}}G_{\rm {F}}(x,x') = -{\rm {i}}G_{{\rm {H}}}(x,x')
    + W(x,x').
\end{equation}
As a solution of the homogeneous scalar field equation, $W(x,x')$ is regular in the limit $x'\rightarrow x$.
Furthermore, $W(x,x')$ depends on the particular quantum state under consideration.
In contrast, $G_{\rm {H}}(x,x')$ is singular in the limit $x'\rightarrow x$ and, since it is a particular solution of the inhomogeneous equation, it is the same for all quantum states.
The renormalization procedure for a charged scalar field in curved space-time is therefore analogous to that described above for a neutral scalar field in flat space-time.
The Feynman Green's function is regularized by subtracting $G_{\rm {H}}(x,x')$ and renormalized expectation values of operators are evaluated by acting with the appropriate differential operators on $W(x,x')$ and then taking the limit $x'\rightarrow x$.

The above procedure depends on the choice of particular solution $G_{\rm {H}}(x,x')$. 
In Minkowski space-time, the particular solution of the inhomogeneous scalar field equation was simply the Feynman Green's function for the vacuum state.
In a general curved space-time, there is no natural choice of vacuum state \cite{Birrell:1982ix} so an alternative method of determining $G_{\rm {H}}(x,x')$ is required.
The choice of $G_{\rm {H}}(x,x')$ must yield physically sensible results for the renormalized expectation values of observables.
A set of physically-motivated axioms which must be satisfied by the renormalized stress-energy tensor was developed by Wald \cite{Wald:1995,Wald:1977up}.
We therefore require $G_{\rm {H}}(x,x')$ to be such that the renormalized stress-energy tensor satisfies Wald's axioms.
It has been rigorously established (see for example \cite{Moretti:1998rf,Moretti:1999ez,Moretti:1998rs,Moretti:1999fb,Hollands:2001nf,Hollands:2001fb,Moretti:2001qh,Hollands:2002ux,Hollands:2004yh} and the references given in \cite{Decanini:2005eg}) that 
the singular part of the Hadamard representation of $G_{\rm {F}}(x,x')$ yields a suitable $G_{\rm {H}}(x,x')$ which, via the above subtraction procedure (known in this case as {\em {Hadamard renormalization}}), leads to a renormalized stress-energy tensor satisfying Wald's axioms. 
We therefore devote the rest of this paper to deriving the form of the Hadamard parametrix $G_{\rm {H}}(x,x')$ for a charged scalar field on a five-dimensional curved space-time.
We first review some geometric quantities which arise in the Hadamard parametrix.

\subsection{Geometric quantities in the Hadamard parametrix}
\label{sec:geometric}

In order to regularize the Feynman Green's function $G_{\rm {F}}(x,x')$, we only require the short-distance behaviour of the singular Hadamard parametrix $G_{\rm {H}}(x,x')$.
We therefore assume that the space-time point $x'$ lies in a normal neighbourhood of the point $x$, so that there is a unique geodesic connecting the two points.
The geodesic distance $\sigma (x,x')$ between $x$ and $x'$ is then well-defined.
It satisfies the curved-space generalization of equation (\ref{eq:sigmaflat}):
\begin{equation}
    2\sigma = g_{\mu \nu }\sigma ^{;\mu }\sigma ^{;\nu },
    \label{eq:sigma}
\end{equation}
where $\sigma ^{;\mu } = \nabla ^{\mu }\sigma $ and $g_{\mu \nu }$ is the curved space-time metric.
From (\ref{eq:Gcloseflat}), it is anticipated that the short-distance behaviour of $G_{\rm {H}}(x,x')$ will depend on $\sigma (x,x')$ (but not necessarily just on $\sigma (x,x')$ as in flat space-time). 
Since $G_{\rm {H}}(x,x')$ satisfies the inhomogeneous scalar field equation (\ref{eq:inhom}), we will need to apply the curved space-time Laplacian $\nabla ^{\mu }\nabla _{\mu }$ to $\sigma (x,x')$.
In five-dimensional Minkowski space-time, this yields simply
\begin{equation}
    \partial ^{\mu }\partial _{\mu }\sigma _{\rm {M}}= \delta ^{\mu }_{\mu } = 5, 
\end{equation}
however in five-dimensional curved space-time we have 
\begin{equation}
    \nabla ^{\mu }\nabla _{\mu } \sigma = 5 - 2\Delta ^{-\frac{1}{2}}\Delta ^{\frac{1}{2}}_{;\mu }\sigma ^{;\mu }.
    \label{eq:boxsigma}
\end{equation}
Here the biscalar $\Delta (x,x')$ is the Van Vleck-Morette determinant \cite{DeWitt1965,Visser:1992pz}
\begin{equation}
    \Delta (x,x') = - \left[  -g(x)\right] ^{-\frac{1}{2}} \det 
    \left[ - \sigma _{;\mu \nu '}(x,x') \right] \left[  -g(x')\right] ^{-\frac{1}{2}},
    \label{eq:VVD}
\end{equation}
where a subscript $;\nu '$ denotes the space-time covariant derivative $\nabla _{\nu '}$ with respect to the space-time point $x'$.
In Minkowski space-time, using (\ref{eq:sigmaflat}), it is straightforward to show that $\Delta _{\rm {M}}(x,x')\equiv 1$ for all $x$, $x'$. 
In a curved space-time, the coincidence limit $x'\rightarrow x$  of geometric quantities can be derived by considering normal coordinates at the fixed space-time point $x$.
Then the leading-order behaviour of both $\sigma (x,x')$ and $g(x)$, $g(x')$ in this coordinate system is the same as in Minkowski space-time, thus
\begin{equation}
    \Delta (x,x)=1.
    \label{eq:Deltabc}
\end{equation}
For a general curved space-time $\Delta (x,x')$ is not identically unity.
If $\Delta (x,x')<1$, a congruence of geodesics emanating from the space-time point $x$ is expanding, while $\Delta (x,x')>1$ indicates that the congruence is focussing \cite{Visser:1992pz,Poisson:2011nh}.

In Minkowski space-time, the singular part of the Feynman Green's function for a neutral scalar field can be written in closed form using a Hankel function (\ref{eq:GFflat}), and has a short-distance expansion which depends only on $\sigma _{\rm {M}}$ (\ref{eq:Gcloseflat}). 
Therefore the singular part of the Feynman Green's function for a neutral scalar field depends only on the distance between the space-time points $x$ and $x'$. 
This is due to the maximal symmetry of Minkowski space-time.
In anti-de Sitter space-time, which is a curved space-time with maximal symmetry, the singular part of the Feynman Green's function for a neutral scalar field also depends only on the geodesic distance $\sigma (x,x')$ between the points \cite{Kent:2014nya}.
However, in a general curved space-time, this will not be the case, and $G_{\rm {H}}(x,x')$ will depend on the direction in which the points are separated as well as the geodesic distance between them.
Even in Minkowski space-time, if we consider a charged scalar field, the presence of the background electromagnetic potential $A_{\mu }$ breaks the maximal symmetry and the singular part of the Feynman Green's function will be direction-dependent.

To take this into account, if we were considering a particular given space-time, one could choose a coordinate patch and expand $G_{\rm {H}}(x,x')$ in terms of the coordinate separation of the points.
However, here we are seeking to develop a general formalism and therefore we will derive a covariant series expansion of $G_{\rm {H}}(x,x')$, which can then be implemented on a given space-time using appropriate coordinates.
The covariant series expansion of a general biscalar $K(x,x')$ involves the tangent vector $\sigma ^{;\mu }$ to the geodesic connecting $x$ to $x'$:
\begin{equation}
    K(x,x') = k_{0}(x)+k_{1\mu }(x)\sigma ^{;\mu } + k_{2\mu \nu }(x)\sigma ^{;\mu }\sigma ^{;\nu } + k_{3\mu \nu \lambda }(x)\sigma ^{;\mu }\sigma ^{;\nu } \sigma ^{;\lambda }+ \ldots , 
\end{equation}
where the coefficients $k_{0}$, $k_{1\mu }$, $k_{2\mu \nu }$, $k_{3\mu \nu \lambda }$, etc.~depend only on the space-time point $x$.
In a particular coordinate system on a given space-time, the tangent vector $\sigma ^{;\mu }$ can be written in terms of the coordinate separation of the points, to give a Taylor series expansion of $K(x,x')$ in that coordinate system (see for example the expansions on a four-dimensional black hole space-time in Appendix B of \cite{Anderson:1994hg}).
From (\ref{eq:sigma}), we can regard $\sigma ^{;\mu }$ as being ${\mathcal {O}}(\sigma ^{\frac{1}{2}})$ and hence, in five-dimensional space-time, we shall ultimately require the covariant series expansion of $G_{\rm {H}}(x,x')$ up to and including terms of the form  $\sigma ^{;\alpha _{1}} \sigma ^{;\alpha _{2}} \sigma ^{;\alpha _{3}} \sigma ^{;\alpha _{4}}\sigma ^{;\alpha _{5}}$.

In our calculations in section~\ref{sec:Ucalc}, we will require the covariant Taylor series expansions of the quantities $\sigma _{;\mu \nu }$, the square root of the Van Vleck-Morette determinant $\Delta  ^{\frac{1}{2}}$, the quantity $\Delta ^{-\frac{1}{2}} \Delta ^{\frac{1}{2}}_{;\mu } \sigma ^{;\mu }$ and $\Box \Delta ^{\frac{1}{2}} = \nabla _{\mu }\nabla ^{\mu } \Delta ^{\frac{1}{2}}$, all of which can be found to high order in \cite{Decanini:2005gt}. 
To the order we require, the expansion for 
$\sigma _{;\mu \nu }$ is
\begin{subequations}
\label{eq:geometricexpansions}
\begin{eqnarray}
 \sigma _{;\mu \nu } & = & 
 g_{\mu \nu } - \frac{1}{3} R_{\mu \alpha _{1} \nu \alpha _{2}} \sigma ^{;\alpha _{1}} \sigma ^{;\alpha _{2}} 
 + \frac{1}{12} R_{\mu \alpha _{1}\nu \alpha _{2} ; \alpha _{3}} 
 \sigma^{;\alpha _{1}} \sigma ^{;\alpha _{2}} \sigma ^{;\alpha _{3}}
 \nonumber \\ & & 
 - \left[ 
 \frac{1}{60} R_{\mu \alpha _{1}\nu \alpha _{2} ; \alpha _{3}\alpha _{4}} +
 \frac{1}{45} R_{\mu \alpha _{1} \rho \alpha _{2}} R^{\rho }{}_{\alpha _{3} \nu \alpha _{4}} 
 \right] \sigma ^{;\alpha _{1}} \sigma ^{;\alpha _{2}} \sigma ^{;\alpha _{3}} \sigma ^{;\alpha _{4}}
 \nonumber \\ & & 
 + \left[ \frac{1}{360} R_{\mu \alpha _{1}\nu \alpha _{2};\alpha _{3}\alpha _{4}\alpha _{5}}
 + \frac{1}{120} R_{\mu \alpha _{1} \rho \alpha _{2}} R^{\rho }{}_{\alpha _{3} \nu \alpha _{4};\alpha _{5}} 
\right. \nonumber \\ & & \left.
\qquad + \frac{1}{120} R_{\mu \alpha _{1} \rho \alpha _{2};\alpha _{3}} R^{\rho }{}_{\alpha _{4}\nu \alpha _{5}} 
 \right]
 \sigma ^{;\alpha _{1}} \sigma ^{;\alpha _{2}} \sigma ^{;\alpha _{3}} \sigma ^{;\alpha _{4}}\sigma ^{;\alpha _{5}}
 + \ldots ,
 \label{eq:sigmaexpansion}
\end{eqnarray}
while that for $\Delta ^{\frac{1}{2}}$ is
\begin{eqnarray}
 \Delta ^{\frac{1}{2}} & = & 
 1 + \frac{1}{12}R_{\alpha _{1}\alpha _{2}} \sigma ^{;\alpha _{1}} \sigma ^{;\alpha _{2}} 
 - \frac{1}{24} R_{\alpha _{1}\alpha _{2};\alpha _{3}} \sigma ^{;\alpha _{1}}\sigma ^{;\alpha _{2}} \sigma ^{;\alpha _{3}}
 \nonumber \\ & & 
 + \left[
 \frac{1}{80} R_{\alpha _{1}\alpha _{2};\alpha _{3}\alpha _{4}} 
 + \frac{1}{360} R^{\rho }{}_{\alpha _{1}\tau \alpha _{2}} R^{\tau }{}_{\alpha _{3}\rho \alpha _{4}} 
 \right. \nonumber \\ & & \left. 
\qquad  + \frac{1}{288} R_{\alpha _{1}\alpha _{2}} R_{\alpha _{3}\alpha _{4}} 
 \right] \sigma ^{;\alpha _{1}} \sigma ^{;\alpha _{2}} \sigma ^{;\alpha _{3}} \sigma ^{;\alpha _{4}}
 \nonumber \\ & & 
 -\left[
 \frac{1}{360} R_{\alpha _{1}\alpha _{2}; \alpha _{3}\alpha _{4}\alpha _{5}}
 +\frac{1}{360}R^{\rho}{}_{\alpha _{1}\tau \alpha _{2}} R^{\tau }{}_{\alpha _{3}\rho \alpha _{4};\alpha _{5}}
 \right. \nonumber \\ & & \left.
\qquad  +\frac{1}{288} R_{\alpha _{1}\alpha _{2}} R_{\alpha _{3}\alpha _{4};\alpha _{5}}
 \right]
 \sigma ^{;\alpha _{1}} \sigma ^{;\alpha _{2}} \sigma ^{;\alpha _{3}} \sigma ^{;\alpha _{4}}\sigma ^{;\alpha _{5}}
 + \ldots ,
 \label{eq:Deltahalf}
\end{eqnarray}
for $\Delta ^{-\frac{1}{2}} \Delta ^{\frac{1}{2}}_{;\mu } \sigma ^{;\mu }$
we have
\begin{eqnarray}
 \Delta ^{-\frac{1}{2}} \Delta ^{\frac{1}{2}}_{;\mu } \sigma ^{;\mu }
 & = & 
 \frac{1}{6} R_{\alpha _{1}\alpha _{2}} \sigma ^{;\alpha _{1}} \sigma ^{;\alpha _{2}} 
 - \frac{1}{24}R_{\alpha _{1}\alpha _{2};\alpha _{3}} \sigma ^{;\alpha _{1}}\sigma ^{;\alpha _{2}} \sigma ^{;\alpha _{3}}
 \nonumber \\ & & 
 + \left[  \frac{1}{120} R_{\alpha _{1}\alpha _{2};\alpha _{3}\alpha _{4}} + \frac{1}{90} R^{\rho }{}_{\alpha _{1}\tau \alpha _{2}} R^{\tau }{}_{\alpha _{3}\rho \alpha _{4}} 
 \right]  \sigma ^{;\alpha _{1}} \sigma ^{;\alpha _{2}} \sigma ^{;\alpha _{3}} \sigma ^{;\alpha _{4}}
 \nonumber \\ & &  
 -\left[ 
 \frac{1}{720} R_{\alpha _{1}\alpha _{2};\alpha _{3}\alpha _{4}\alpha _{5}}
 + \frac{1}{120} R^{\rho} {}_{\alpha _{1}\tau \alpha _{2}} R^{\tau}{}_{\alpha _{3}\rho \alpha _{4};\alpha _{5}} 
 \right]
 \sigma ^{;\alpha _{1}} \sigma ^{;\alpha _{2}} \sigma ^{;\alpha _{3}} \sigma ^{;\alpha _{4}}\sigma ^{;\alpha _{5}}
 \nonumber \\ & & 
 + \ldots ,
 \label{eq:Dderivexpansion}
\end{eqnarray}
and finally, for $\Box \Delta ^{\frac{1}{2}}$ we only require the expansion to ${\mathcal {O}}(\sigma ^{\frac{3}{2}})$, which is
\begin{eqnarray}
\Box \Delta ^{\frac{1}{2}}
& = & 
\frac{1}{6}R 
+ \left[ 
\frac{1}{40} \Box R_{\alpha _{1}\alpha _{2}}
- \frac{1}{120} R_{;\alpha _{1}\alpha _{2}}
+ \frac{1}{72} R R_{\alpha _{1}\alpha _{2}}
- \frac{1}{30} R^{\rho }{}_{\alpha _{1}}R_{\rho \alpha _{2}}
\right. \nonumber \\ & & \left.
\qquad + \frac{1}{60} R^{\rho \tau }R_{\rho \alpha _{1}\tau \alpha _{2}}
+ \frac{1}{60} R^{\rho \kappa \tau }{}_{\alpha _{1}}R_{\rho \kappa \tau \alpha _{2}}
\right] \sigma ^{;\alpha _{1}}\sigma ^{;\alpha _{2}}
\nonumber \\ & & 
+\left[ 
\frac{1}{360} R_{;\alpha _{1}\alpha _{2}\alpha _{3}}
- \frac{1}{120} \left( \Box R_{\alpha _{1}\alpha _{2}} \right) _{;\alpha _{3}}
- \frac{1}{144} R R_{\alpha _{1}\alpha _{2};\alpha _{3}}
+ \frac{1}{45} R^{\rho }{}_{\alpha _{1}}R_{\rho \alpha _{2};\alpha _{3}} 
\right. \nonumber \\ & & \left. 
\qquad - \frac{1}{180} R^{\rho }{}_{\tau ; \alpha _{1}}R^{\tau }{}_{\alpha _{2}\rho \alpha _{3}}
- \frac{1}{180} R^{\rho }{}_{\tau } R^{\tau }{}_{\alpha _{1}\rho \alpha _{2}; \alpha _{3}}
\right. \nonumber \\ & & \left. 
\qquad - \frac{1}{90} R^{\rho \kappa \tau }{}_{\alpha _{1}}R_{\rho \kappa \tau \alpha _{2};\alpha _{3}}
\right] \sigma ^{;\alpha _{1}}\sigma ^{;\alpha _{2}}\sigma ^{;\alpha _{3}}
+\ldots .
\label{eq:boxDelta}
\end{eqnarray}
We also require the expansion of $\Delta ^{\frac{1}{2}}_{;\mu }$ to ${\mathcal {O}}(\sigma ^2)$.
This can be found by differentiating (\ref{eq:Deltahalf}) and simplifying using (\ref{eq:sigmaexpansion}):
\begin{eqnarray}
\Delta ^{\frac{1}{2}}_{;\mu } & = & 
\frac{1}{6} R_{\mu \alpha _{1}}\sigma ^{;\alpha _{1}}
+ \left[ \frac{1}{24} R_{\alpha _{1}\alpha _{2};\mu } - \frac{1}{12} R_{\mu (\alpha _{1};\alpha _{2})} \right]  \sigma ^{;\alpha _{1}}\sigma ^{;\alpha _{2}}
\nonumber \\ & & 
+\left[ 
\frac{1}{40} R_{\mu (\alpha _{1}; \alpha _{2}\alpha _{3})}
-\frac{1}{60} R_{(\alpha _{1}\alpha _{2};|\mu |\alpha _{3})}
+\frac{1}{72} R_{\mu (\alpha _{1}}R_{\alpha _{2}\alpha _{3})}
\right. \nonumber \\ & & \left. 
\qquad + \frac{1}{90} R^{\rho }{}_{\mu  \tau (\alpha _{1} }R^{\tau }{}_{\alpha _{2}|\rho | \alpha _{3})}
+ \frac{1}{360} R_{\rho (\alpha _{1}}R^{\rho }{}_{\alpha _{2}|\mu |\alpha _{3})}
\right]  \sigma ^{;\alpha _{1}}\sigma ^{;\alpha _{2}} \sigma ^{;\alpha _{3}} 
\nonumber \\ & & 
+
\left[
\frac{1}{240}  R_{(\alpha _{1} \alpha _{2} ;  \alpha _{3}\alpha _{4}) \mu }
-\frac{1}{180} R_{\mu (\alpha _{1};\alpha _{2}\alpha _{3}\alpha _{4})}
+ \frac{1}{288} R_{(\alpha _{1}\alpha _{2}}R_{\alpha _{3}\alpha _{4});\mu }
\nonumber
\right. \\ & & \left. 
\qquad -\frac{1}{144} R_{\mu (\alpha _{1};\alpha _{2}} R_{ \alpha _{3}\alpha _{4})}
- \frac{1}{144} R_{\mu ( \alpha _{1}}R_{\alpha _{2}\alpha _{3};\alpha _{4})}
+ \frac{1}{90} R_{\rho (\alpha _{1};\alpha _{2}} R^{\rho }{}_{\alpha _{3} | \mu | \alpha _{4})}
\nonumber
\right. \\ & & \left. 
\qquad 
+ \frac{1}{120} R^{\rho }{}_{(\alpha _{1} | \mu | \alpha _{2}}R_{\alpha _{3}\alpha _{4});\rho }
+ \frac{1}{120} R_{\rho (\alpha _{1}} R^{\rho }{}_{\alpha _{2}|\mu | \alpha _{3};\alpha _{4})}
\nonumber
\right. \\ & & \left. 
\qquad 
+ \frac{1}{360} R^{\rho }{}_{(\alpha _{1}| \tau | \alpha _{2}}
R^{\tau }{}_{\alpha _{3}|\rho | \alpha _{4});\mu }
-\frac{1}{360} R^{\rho }{}_{\mu \tau (\alpha _{1}}
R^{\tau }{}_{\alpha _{2} | \rho | \alpha _{3};\alpha _{4})}
\nonumber
\right. \\ & & \left. 
\qquad 
- \frac{1}{360} R^{\rho }{}_{(\alpha _{1} | \tau | \alpha _{2}}
R^{\tau }{}_{|\mu \rho | \alpha _{3};\alpha _{4})}
- \frac{1}{360} R^{\rho }{}_{(\alpha _{1} |\tau | \alpha _{2}}
R^{\tau }{}_{\alpha _{3}|\rho \mu | ;\alpha _{4})}
\nonumber
\right. \\ & & \left. 
\qquad 
-\frac{1}{360} R^{\rho }{}_{(\alpha _{1}|\tau \mu |}
R^{\tau }{}_{\alpha _{2}|\rho | \alpha _{3};\alpha _{4})}
\right] \sigma ^{;\alpha _{1}}\sigma ^{;\alpha _{2}} \sigma ^{;\alpha _{3}}  \sigma ^{;\alpha _{4}}
+ \ldots .
\label{eq:DeltaDexpansion}
\end{eqnarray}
\end{subequations}
The notation $( \ldots  )$ denotes symmetrization over the indices enclosed in the brackets, with vertical bars $| \rho |$ surrounding indices $\rho $ which are not included in the symmetrization.
In the covariant series expansions (\ref{eq:sigmaexpansion}--\ref{eq:Deltahalf}), all terms except the zeroth order ones depend on curvature tensors and hence vanish in flat space-time.
Therefore, in Minkowski space-time, we have $\sigma _{;\mu \nu }=\eta _{\mu \nu }$ and $\Delta (x,x')=1$, as expected.
In addition, $\Box \Delta ^{\frac{1}{2}}$ (\ref{eq:boxDelta}) and $\Delta _{;\mu }^{\frac{1}{2}}$ (\ref{eq:DeltaDexpansion}) depend only on curvature tensors and so vanish in Minkowski space-time.
We emphasize that the quantities $\sigma (x,x')$ and $\Delta (x,x')$ are purely geometric and independent of the scalar field parameters $q$, $m$ and $\xi $. 
In particular, the expansions (\ref{eq:geometricexpansions}) are valid whether we are considering a neutral or charged scalar field.

\subsection{Hadamard parametrix for the singular part of the Feynman Green's function}

For a charged scalar field on a five-dimensional curved space-time, the singular part of the Hadamard representation of the Feynman Green's function takes the form \cite{Balakumar:2019djw}
\begin{equation}
    -{\rm {i}}G_{\rm {H}}(x,x') = 
    \frac{1}{16{\sqrt {2}}\pi ^{2}}
    \frac{U(x,x')}{\sigma (x,x') ^{\frac{3}{2}}} ,
    \label{eq:GFcurved}
\end{equation}
where we have assumed that $\sigma (x,x') >0$, omitted the $\epsilon \rightarrow 0$, and the biscalar $U(x,x')$ is to be determined using the fact that $G_{\rm {H}}(x,x')$ is a solution of the inhomogeneous scalar field equation (\ref{eq:inhom}).
The parametrix (\ref{eq:GFcurved}) has the same general form as that for a neutral scalar field in five dimensions \cite{Decanini:2005eg}, but the biscalar $U(x,x')$ will be different for a charged scalar field compared with the neutral case.

First we consider the expression (\ref{eq:GFcurved}) in normal coordinates at the fixed space-time point $x$.
For $x'$ sufficiently close to $x$, in this coordinate system the principal part of the second order differential operator appearing in (\ref{eq:inhom}) reduces to $\partial _{\mu }\partial ^{\mu }$ plus terms proportional to the square of the coordinate separation of the points.
Therefore, to leading order in the coordinate separation, the singular part of the Feynman Green's function (\ref{eq:GFcurved}) must match the leading  order divergence of the Minkowski space-time Feynman Green's function (\ref{eq:Gcloseflat}), so that
\begin{equation}
    \lim _{x'\rightarrow x}U(x,x') =1. 
    \label{eq:Uboundary}
\end{equation}
In particular, the biscalar $U(x,x')$ is regular in the coincidence limit.

The differential equation satisfied by $U(x,x')$ is derived by substituting (\ref{eq:GFcurved}) into (\ref{eq:inhom}), assuming that $\sigma (x,x')>0$,
which gives
\begin{eqnarray}
 0 & = & 
 \frac{1}{\sigma ^{\frac{3}{2}}}\left\{ 
 \left[ D_{\mu }D^{\mu }- m^{2}-\xi R\right] U 
 - \frac{3}{\sigma } (D_{\mu }U)\sigma ^{;\mu }
 -\frac{3}{2\sigma }U \nabla _{\mu }\nabla ^{\mu }\sigma 
 +\frac{15}{4\sigma ^{2}}U\sigma _{;\mu }\sigma ^{;\mu }
 \right\} 
 \nonumber \\ & = & 
  \frac{1}{\sigma ^{\frac{3}{2}}}\left\{ 
 \left[ D_{\mu }D^{\mu }- m^{2}-\xi R\right] U 
 - \frac{3}{\sigma } (D_{\mu }U)\sigma ^{;\mu }
+\frac{3U}{\sigma } \Delta ^{-\frac{1}{2}}\Delta ^{\frac{1}{2}}_{;\mu }\sigma ^{;\mu } 
 \right\} ,
\end{eqnarray}
where in the second line we have simplified using (\ref{eq:sigma}, \ref{eq:boxsigma}).
Therefore the biscalar $U(x,x')$ satisfies the homogeneous differential equation \cite{Balakumar:2019djw}
\begin{equation}
  \sigma \left[ D_{\mu }D^{\mu }- m^{2}-\xi R\right] U 
 - 3 \sigma ^{;\mu } D_{\mu }U
+3U \Delta ^{-\frac{1}{2}}\Delta ^{\frac{1}{2}}_{;\mu }\sigma ^{;\mu }   =0. 
\label{eq:Ueqn}
\end{equation}
Following \cite{Decanini:2005gt,Hadamard,Garabedian,Friedlander} 
we now write $U(x,x')$ as a power series in $\sigma (x,x')$ as follows:
\begin{equation}
    U(x,x') = \sum _{n=0}^{\infty } U_{n}(x,x') \sigma ^{n}(x,x'),
    \label{eq:Upower}
\end{equation}
where the biscalars $U_{n}(x,x')$ are regular in the coincidence limit $x'\rightarrow x$ and 
\begin{equation}
    U_{0}(x,x)=1 .
    \label{eq:U0boundary}
\end{equation}
Substituting (\ref{eq:Upower}) into (\ref{eq:Ueqn}), we find
\begin{eqnarray}
 0 & = & \sum _{n=0}^{\infty }\left\{
 \sigma \left[ D_{\mu }D^{\mu }- m^{2}-\xi R\right] U_{n} 
 + \left( 2n -3 \right) \sigma ^{;\mu }D_{\mu }U_{n}
 + n U_{n}\nabla _{\mu }\nabla ^{\mu }\sigma
 \right. \nonumber \\ & & \left.
 + n(n-4) U_{n}\sigma ^{-1} \sigma _{;\mu}\sigma ^{;\mu }
 +3U_{n}\Delta ^{-\frac{1}{2}}\Delta ^{\frac{1}{2}}_{;\mu }\sigma ^{;\mu }
 \right\} \sigma ^{n}
 \nonumber \\ & = & 
 \sum _{n=0}^{\infty }\left\{
 \sigma \left[ D_{\mu }D^{\mu }- m^{2}-\xi R\right] U_{n}
 + \left( 2n -3 \right) \sigma ^{;\mu }D_{\mu }U_{n}
 + n(2n-3) U_{n}
 \right. \nonumber \\ & & \left.
 +(3-2n)U_{n}\Delta ^{-\frac{1}{2}}\Delta ^{\frac{1}{2}}_{;\mu }\sigma ^{;\mu }
 \right\} \sigma ^{n} ,
\end{eqnarray}
again using (\ref{eq:sigma}, \ref{eq:boxsigma}).
The zeroth order term gives the equation satisfied by $U_{0}(x,x')$, namely
\cite{Balakumar:2019djw}
\begin{subequations}
\label{eq:Ueqnsgeneral}
\begin{equation}
\sigma ^{;\mu }D_{\mu }U_{0} - \Delta ^{-\frac{1}{2}}\Delta ^{\frac{1}{2}}_{;\mu }\sigma ^{;\mu } U_{0}=0,
    \label{eq:U0eqn}
\end{equation}
while setting the coefficient of $\sigma ^{n}$ for $n>0$ to vanish gives
\begin{equation}
  \left[ D_{\mu }D^{\mu }- m^{2}-\xi R\right] U_{n-1}
  + \left( 2n -3 \right)
  \left[ \sigma ^{;\mu }D_{\mu }
 +n -\Delta ^{-\frac{1}{2}}\Delta ^{\frac{1}{2}}_{;\mu }\sigma ^{;\mu } \right] U_{n} =0 .
 \label{eq:Uneqn}
\end{equation}
\end{subequations}
If we consider a neutral scalar field, the covariant derivatives $D_{\mu }$ in (\ref{eq:Ueqnsgeneral}) reduce to $\nabla _{\mu }$, yielding the equations for $U_{n}$ given in \cite{Decanini:2005eg}.
In particular, the equation (\ref{eq:U0eqn}) for $U_{0}(x,x')$ takes the form
\begin{equation}
\sigma ^{;\mu }\nabla _{\mu }U_{0} - \Delta ^{-\frac{1}{2}}\Delta ^{\frac{1}{2}}_{;\mu }\sigma ^{;\mu } U_{0}=0,
    \label{eq:U0eqnneutral}
\end{equation}
which has the solution $U_{0}(x,x')=\Delta ^{\frac{1}{2}}(x,x')$ for a neutral scalar field.
For a charged scalar field, all the coefficients $U_{n}(x,x')$ (including $U_{0}(x,x')$) will depend on the electromagnetic potential $A_{\mu }$ as well as geometric quantities.

\section{Covariant series expansion of the Hadamard coefficients}
\label{sec:Ucalc}

In this section we consider in detail the covariant series expansions of the Hadamard coefficients $U_{n}(x,x')$.
Since the leading order divergence of the Hadamard parametrix (\ref{eq:GFcurved}) is ${\mathcal {O}}(\sigma ^{-\frac{3}{2}})$,
the computation of the renormalized scalar field condensate requires $U(x,x')$ to be known up to order ${\mathcal {O}}(\sigma ^{\frac{3}{2}})$.
Finding the renormalized current (\ref{eq:currentdef}) involves taking one derivative of the Feynman Green's function and hence requires knowledge of $U(x,x')$ up to ${\mathcal {O}}(\sigma ^{2})$, while the renormalized stress-energy tensor (\ref{eq:Tmunudef}) involves two derivatives and therefore $U(x,x')$ to order ${\mathcal {O}}(\sigma ^{\frac{5}{2}})$.
We thus consider the following expansions:
\begin{subequations}
\label{eq:Uexpansions}
\begin{eqnarray}
 U_{0}(x,x') & = & 
 U_{00} + U_{01\mu }\sigma ^{;\mu }+ U_{02\mu \nu }\sigma ^{;\mu }\sigma ^{;\nu } 
 +U_{03\mu \nu \lambda } \sigma ^{;\mu }\sigma ^{;\nu }\sigma ^{;\lambda }
 + U_{04\mu \nu \lambda \tau } \sigma ^{;\mu }\sigma ^{;\nu }\sigma ^{;\lambda }\sigma ^{;\tau }
 \nonumber \\ & & 
 + U_{05\mu \nu \lambda \tau \rho } \sigma ^{;\mu }\sigma ^{;\nu }\sigma ^{;\lambda }\sigma ^{;\tau }\sigma ^{;\rho } + \ldots 
 \label{eq:U0expansion}
 \\
 U_{1}(x,x') & = & 
 U_{10} + U_{11\mu }\sigma ^{;\mu }+ U_{12\mu \nu }\sigma ^{;\mu }\sigma ^{;\nu } 
 +U_{13\mu \nu \lambda } \sigma ^{;\mu }\sigma ^{;\nu }\sigma ^{;\lambda }
 + \ldots 
 \label{eq:U1expansion}
 \\
 U_{2}(x,x') & = & 
 U_{20} + U_{21\mu }\sigma ^{;\mu }
 + \ldots 
 \label{eq:U2expansion}
\end{eqnarray}
\end{subequations}
The coefficients $U_{00}$, $U_{01\mu }$, etc.~all depend only on the space-time point $x$ and are symmetric tensors.
To find these coefficients, we substitute the expansions (\ref{eq:Uexpansions}) into the equations (\ref{eq:Ueqnsgeneral}) and compare term-by-term.
In this section we include more steps in the derivation than presented in our earlier work \cite{Balakumar:2019djw}, since these may be useful
for researchers wishing to perform similar calculations. 

The coefficients in the expansions (\ref{eq:Uexpansions}) typically contain three types of terms: 
\begin{enumerate}
    \item expressions involving the metric and curvature tensors only, 
    \label{list:curvature}
    \item expressions involving only the electromagnetic potential and gauge field strength and their derivatives,
    \label{list:EM}
 \item expressions involving both curvature tensors and gauge field quantities.
 \label{list:both}
\end{enumerate}
In the absence of the scalar field charge, only terms of type (\ref{list:curvature}) remain and these  can be found in \cite{Decanini:2005eg}. 
In Minkowski space-time, most terms of type (\ref{list:curvature}) and all terms of type (\ref{list:both}) vanish identically, leaving just a few quantities depending on the scalar field mass $m$ and terms of type (\ref{list:EM}). 

A major challenge in the calculations whose results we report in this section is simplifying the expressions to give as compact a form as possible. 
Our general simplification strategy is to combine similar terms as far as possible, if necessary changing the order in which covariant derivatives are taken, and exploiting the symmetry of the tensors. 
To this end, we first note that the electromagnetic field strength $F_{\mu \nu }$ is an antisymmetric tensor given by
\begin{equation}
    F_{\mu \nu }= D_{\mu }A_{\nu }-  D_{\nu }A_{\mu },
\end{equation}
and we will find the following identities (and their derivatives) to be very useful:
\begin{subequations}
\label{eq:identities}
\begin{eqnarray}
\left[ D_{\mu }D_{\nu } - D_{\nu }D_{\mu } \right] A_{\rho }
& = & 
R^{\alpha }{}_{\rho \nu \mu }A_{\alpha }- {\rm {i}}q A_{\rho }F_{\mu \nu },
\\
\left[ 
D_{\mu }D_{\nu }D_{\lambda } - D_{\nu }D_{\mu } D_{\lambda }
\right] A_{\rho }
& = &
R^{\alpha }{}_{\lambda \nu \mu }D_{\alpha }A_{\rho }
+ R^{\alpha }{}_{\rho \nu \mu }D_{\lambda }A_{\alpha }
-{\rm {i}}q F_{\mu \nu }D_{\lambda }A_{\rho }.
\nonumber \\ & & 
 \end{eqnarray}
 \end{subequations}
 Due to the antisymmetry of $F_{\mu \nu }$, and the symmetry of the Ricci tensor $R_{\mu \nu }$, we also have the identity
 \begin{equation}
     \nabla ^{\mu }\nabla ^{\nu }F_{\mu \nu }=0.
 \end{equation}
It is worth noting that the covariant derivatives $D_{\mu }$ do not commute even in Minkowski space-time, due to the electromagnetic field.
From the identities (\ref{eq:identities}), it can be seen that changing the orders of derivatives invariably introduces terms of type (\ref{list:both}). 
The number of such terms increases rapidly as the order of the expansion increases and the number of derivatives that need to be considered increases.
We find that the highest order terms in the expansions (\ref{eq:Uexpansions}) of $U_{1}$ and $U_{2}$, namely $U_{13\mu \nu \lambda }$ and $U_{21\mu }$, contain many terms of type (\ref{list:both}). 
Since these lengthy expressions are not particularly informative, for reasons of space we present expressions for $U_{13\mu \nu \lambda }$ and $U_{21\mu }$ only in Minkowski space-time, but for a general background electromagnetic potential $A_{\mu }$.
The expressions we give for all other terms in the expansions (\ref{eq:Uexpansions}) will be valid for arbitrary $A_{\mu }$ and space-time metric $g_{\mu \nu }$.

\subsection{$U_{0}$}
\label{sec:U0}

First we find $U_{0}$ by solving (\ref{eq:U0eqn}).
This equation is identical to that satisfied by $U_{0}$ in any number of space-time dimensions greater than or equal to three \cite{Balakumar:2019djw}.
In \cite{Balakumar:2019djw}, this equation was solved up to order ${\mathcal {O}}(\sigma ^{2})$ in four space-time dimensions, but since we are working here in five dimensions, we require $U_{0}$ to order ${\mathcal {O}}(\sigma ^{\frac{5}{2}})$.
In this section we provide more details of the calculation in \cite{Balakumar:2019djw}, as well as extending it to the required higher order.

The equation (\ref{eq:U0eqn}) involves only one derivative of $U_{0}$, which is given by 
\begin{eqnarray}
 D_{\alpha  }U_{0} & = & 
 D_{\alpha  }U_{00} + \left( D_{\alpha  }U_{01\mu }\right)\sigma ^{;\mu }
+ g^{\mu \beta} U_{01\mu } \sigma _{;\beta \alpha }
 + \left( D_{\alpha }U_{02\mu \nu } \right) \sigma ^{;\mu }\sigma ^{;\nu }
 \nonumber \\ & &
 + 2 g^{\mu \beta } U_{02 \mu \nu }\sigma _{;\beta \alpha } \sigma ^{;\nu }
 +\left( D_{\alpha }U_{03\mu \nu \lambda }\right) \sigma ^{;\mu }\sigma ^{;\nu }\sigma ^{;\lambda }
 + 3 g^{\mu \beta }U_{03\mu \nu \lambda }\sigma _{;\beta \alpha }\sigma ^{;\nu }\sigma ^{;\lambda }
  \nonumber \\ & & 
 + \left( D_{\alpha }U_{04\mu \nu \lambda \tau }\right) \sigma ^{;\mu }\sigma ^{;\nu }\sigma ^{;\lambda }\sigma ^{;\tau }
 + 4 g^{\mu \beta }U_{04\mu \nu \lambda \tau }\sigma _{;\beta \alpha }
 \sigma ^{;\nu }\sigma ^{;\lambda }\sigma ^{;\tau }
 \nonumber \\ & & 
 +\left( D_{\alpha } U_{05\mu \nu \lambda \tau \rho } \right) \sigma ^{;\mu }\sigma ^{;\nu }\sigma ^{;\lambda }\sigma ^{;\tau }\sigma ^{;\rho } 
 + 5g^{\mu \beta } U_{05\mu \nu \lambda \tau \rho } \sigma _{;\beta \alpha }\sigma ^{;\nu }\sigma ^{;\lambda }\sigma ^{;\tau }\sigma ^{;\rho } 
 + \ldots ,
 \nonumber \\ & & 
 \label{eq:DaU0}
\end{eqnarray}
where we have made use of the fact that the expansion coefficients are symmetric tensors.
This expression can be simplified using (\ref{eq:sigmaexpansion}), and we retain terms up to ${\mathcal {O}}(\sigma ^{2})$:
\begin{eqnarray}
 D_{\alpha }U_{0} & = & 
 \left[U_{01\alpha } + D_{\alpha  }U_{00} \right] 
+ \left[ 2U_{02\alpha \mu } + D_{\alpha  }U_{01\mu } \right] \sigma ^{;\mu }
 \nonumber \\ & &
 + \left[ 3U_{03\alpha \mu \nu }+ D_{\alpha }U_{02\mu \nu } 
 - \frac{1}{3} U_{01\rho } R^{\rho }{}_{ \mu \alpha \nu}
 \right] \sigma ^{;\mu }\sigma ^{;\nu }
 \nonumber \\ & &
 + \left[ 4U_{04\alpha \mu \nu \lambda } + D_{\alpha }U_{03\mu \nu \lambda}  -\frac{2}{3} U_{02 \rho \lambda  } R^{\rho }{}_{ \mu \alpha \nu}
 +\frac{1}{12} U_{01\rho } R^{\rho }{}_{ \mu \alpha \nu ; \lambda }
 \right] \sigma ^{;\mu }\sigma ^{;\nu }\sigma ^{;\lambda }
 \nonumber \\ & &
 + \left[ 5U_{05\alpha \mu \nu \lambda \tau }+D_{\alpha }U_{04\mu \nu \lambda \tau }
 -U_{03 \rho \lambda \tau  } R^{\rho }{}_{ \mu \alpha \nu} 
 +\frac{1}{6} U_{02\rho \tau } R^{\rho }{}_{ \mu \alpha \nu ; \lambda }
 \right. \nonumber \\ & & \left.
 \qquad -  U_{01\rho } \left( \frac{1}{60} R^{\rho }{}_{\mu \alpha \nu ; \lambda \tau }+ \frac{1}{45} R^{\rho }{}_{\mu \kappa \nu }R^{\kappa   }{}_{\lambda \alpha \tau }  \right) 
 \right] \sigma ^{;\mu }\sigma ^{;\nu }\sigma ^{;\lambda }\sigma ^{;\tau }
 + \ldots
 \label{eq:Dalphasimp}
\end{eqnarray}
We now substitute $D_{\alpha }U_{0}$ into (\ref{eq:U0eqn}) and set the coefficients of the resulting covariant series expansion to zero.
When we multiply $D_{\alpha }U_{0}$ by $\sigma ^{;\alpha }$, all the terms in (\ref{eq:Dalphasimp}) involving the Riemann tensor disappear since the latter is antisymmetric in its last two indices.
The other term in (\ref{eq:U0eqn}) involves $\Delta ^{-\frac{1}{2}} \Delta ^{\frac{1}{2}}_{;\mu }\sigma ^{;\mu }$, whose expansion is given in (\ref{eq:Dderivexpansion}).
Multiplying this by the expansion for  $U_{0}$ (\ref{eq:U0expansion}) gives a covariant Taylor series whose lowest term is ${\mathcal {O}}(\sigma )$.
We therefore obtain the following simplified set of equations
\begin{subequations}
\label{eq:U0eqns}
\begin{eqnarray}
 0 & = & 
 U_{01\alpha }+ D_{\alpha }U_{00},
 \label{eq:U01eqn}
 \\
 0 & = & 
  2U_{02\alpha \mu } + D_{(\alpha  }U_{01\mu )}  - \frac{1}{6}R_{\alpha \mu }U_{00} ,
  \label{eq:U02eqn}
  \\
  0 & = & 
  3U_{03\alpha \mu \nu }+ D_{(\alpha }U_{02\mu \nu )} 
  - \frac{1}{6}R_{(\alpha \mu }U_{01 \nu )}
 + \frac{1}{24} R_{(\alpha \mu ;\nu )}U_{00} ,
 \label{eq:U03eqn}
\\
 0 & = & 
 4U_{04\alpha \mu \nu \lambda } + D_{(\alpha }U_{03\mu \nu \lambda )}
 - \frac{1}{6}R_{(\alpha \mu } U_{02\nu \lambda )}
  + \frac{1}{24} R_{(\alpha \mu ;\nu }U_{01 \lambda )} 
   \nonumber \\ & & 
 -\frac{1}{120} R_{(\alpha \mu ; \nu \lambda )}  U_{00}
 - \frac{1}{90} R^{\rho }{}_{(\alpha | \kappa  | \mu }R^{\kappa }{}_{\nu |\rho | \lambda )}U_{00}
 ,
 \label{eq:U04eqn}
\\
 0 & = & 
 5U_{05\alpha \mu \nu \lambda \tau }+D_{(\alpha }U_{04\mu \nu \lambda \tau )}
  - \frac{1}{6}R_{(\alpha \mu } U_{03 \nu \lambda \tau )}
  + \frac{1}{24} R_{(\alpha \mu ;\nu }U_{02 \lambda \tau ) }
  \nonumber \\ & & 
  -\frac{1}{120} R_{(\alpha \mu ; \nu \lambda }  U_{01 \tau )}
 - \frac{1}{90} R^{\rho }{}_{(\alpha | \kappa  | \mu }R^{\kappa }{}_{\nu |\rho | \lambda  }U_{01 \tau)}
 \nonumber \\ & & 
  + 
 \frac{1}{720} R_{(\alpha \mu ;\nu \lambda \tau )}U_{00}
 +\frac{1}{120} R^{\rho} {}_{(\alpha |\kappa |\mu } R^{\kappa }{}_{\nu  | \rho |\lambda ;\tau )}U_{00},
\label{eq:U05eqn}
\end{eqnarray}
\end{subequations}
where we have used the fact that $U_{02\mu \nu }$, $U_{03\mu \nu \lambda }$, \ldots are symmetric tensors.
The terms in the equations (\ref{eq:U0eqns}) involving curvature tensors arise from the expression $\Delta ^{-\frac{1}{2}} \Delta ^{\frac{1}{2}}_{;\mu }\sigma ^{;\mu }U_{0}$ in (\ref{eq:U0eqn}), with the remainder coming from $\sigma ^{;\mu }D_{\mu }U_{0}$. 

The equations (\ref{eq:U0eqns}) can be solved iteratively.
First of all, the boundary condition (\ref{eq:U0boundary}) gives
\begin{subequations}
\label{eq:U0}
\begin{equation}
    U_{00} = 1 .
    \label{eq:U00}
\end{equation}
Substituting this into the first of our set of equations (\ref{eq:U01eqn}) straightforwardly yields
\begin{equation}
    U_{01\mu } = -\nabla _{\mu }U_{00} + {\rm {i}}q A_{\mu }U_{00} = {\rm {i}}q A_{\mu }.
    \label{eq:U01}
\end{equation}
Substituting for $U_{01\mu }$ in (\ref{eq:U02eqn}) then gives
\begin{equation}
    U_{02 \mu \nu } = \frac{1}{12}R_{\mu \nu }U_{00} - \frac{1}{2}D_{(\mu }U_{01\nu )} 
    = \frac{1}{12}R_{\mu \nu } - \frac{{\rm{i}}q}{2}D_{(\mu }A_{\nu )}. 
    \label{eq:U02}
\end{equation}
Proceeding in a similar way, the expressions become increasingly complicated. 
After further simplifications, we find the following results:
\begin{eqnarray}
 U_{03\mu \nu \lambda } & = & 
 -\frac{1}{24} R_{( \mu \nu ; \lambda )}
 + \frac{{\rm {i}}q}{6}D_{(\mu }D_{\nu }A_{\lambda )}
+ \frac{{\rm {i}}q}{12} A_{(\mu }R_{\nu  \lambda )},
\\
 U_{04\mu \nu \lambda \tau } & = & 
 \frac{1}{80} R_{(\mu \nu ; \lambda \tau )} + \frac{1}{360}R^{\rho }{}_{(\mu | \kappa  | \nu }R^{\kappa }{}_{\lambda | \rho | \tau )} 
 + \frac{1}{288} R_{(\mu \nu }R_{\lambda \tau )}
 \nonumber \\ & & 
-\frac{{\rm {i}}q}{24} D_{(\mu }D_{\nu }D_{\lambda }A_{\tau )} - \frac{{\rm {i}}q}{24}D_{(\mu }\left[ A_{\nu }R_{\lambda \tau )} \right] ,
\\
U_{05\mu \nu \lambda \tau \rho} & = & 
- \frac{1}{360} R_{(\mu \nu ; \lambda \tau \rho )}
- \frac{1}{288} R_{(\mu \nu }R_{\lambda \tau ;\rho )}
- \frac{1}{360} R^{\kappa }{}_{(\mu | \psi | \nu }R^{\psi }{}_{\lambda |\kappa  |\tau ; \rho )} 
\nonumber \\ & & 
+ \frac{{\rm {i}}q}{120} D_{(\mu } D_{\nu }D_{\lambda }D_{\tau }A_{\rho )}
+\frac{{\rm {i}}q}{96} D_{(\mu }D_{\nu }\left[ A_{\lambda }R_{\tau \rho )} \right] 
\nonumber \\ & & 
+\frac{{\rm {i}}q}{288} R_{(\mu \nu }D_{\lambda }D_{\tau }A_{\rho )}
+\frac{{\rm {i}}q}{480} A_{(\mu } R_{\nu \lambda ; \tau \rho )} 
+\frac{{\rm {i}}q}{288} A_{(\mu } R_{\nu \lambda }R_{\tau \rho )}
\nonumber \\ & & 
+\frac{{\rm {i}}q}{360} A_{(\mu }  R^{\kappa }{}_{\nu | \psi | \lambda  }R^{\psi }{}_{\tau | \kappa | \rho )} .
\end{eqnarray}
\end{subequations}
It can be seen that the complexity of the expressions increases substantially as the order of the expansion increases.  
We have been able to find the expansion coefficients in $U_{0}$ in a comparatively neat form in terms of the gauge covariant derivative of the electromagnetic potential $A_{\mu }$.
As expected, the first five terms in (\ref{eq:U0}) are identical to those given in \cite{Balakumar:2019djw} in four space-time dimensions.

Since the electromagnetic potential does not affect the principal part of the partial differential operator in the inhomogeneous scalar field equation (\ref{eq:inhom}), it also does not affect the leading order terms in $U(x,x')$, and makes its first appearance at ${\mathcal {O}}(\sigma ^{\frac{1}{2}})$ in $U_{0}$.
We see that $U_{0}$ depends on the space-time curvature and background electromagnetic potential, but does not depend on the scalar field mass $m$ or coupling $\xi $ to the curvature scalar.

If $q=0$, then the expressions (\ref{eq:U0}) reduce to those for a neutral scalar field in five space-time dimensions given in \cite{Decanini:2005eg}.
In this case $U_{0}=\Delta ^{\frac{1}{2}}$, as can be seen by comparing the expansions (\ref{eq:Deltahalf}) and (\ref{eq:U0}). 
For a charged scalar field on a Minkowski space-time background, all terms in (\ref{eq:U0}) which involve curvature tensors vanish identically, and we are left with
\begin{eqnarray}
    U_{0}^{\rm {M}} & = & 1 +{\rm {iq}} \left[ 
    A_{\mu} \sigma ^{;\mu }
    - \frac{1}{2}D_{(\mu } A_{\nu )}\sigma ^{;\mu }\sigma ^{;\nu }
     + \frac{1}{6}D_{(\mu }D_{\nu }A_{\lambda )}\sigma ^{;\mu }\sigma ^{;\nu } \sigma ^{;\lambda }
\right.     \nonumber \\ & &  \left. \hspace{-0.5cm}
     -\frac{1}{24} D_{(\mu }D_{\nu }D_{\lambda }A_{\tau )}
     \sigma ^{;\mu }\sigma ^{;\nu } \sigma ^{;\lambda }\sigma ^{;\tau }
     + \frac{1}{120} D_{(\mu } D_{\nu }D_{\lambda }D_{\tau }A_{\rho )}
     \sigma ^{;\mu }\sigma ^{;\nu } \sigma ^{;\lambda }\sigma ^{;\tau }\sigma ^{;\rho }
     + \ldots \right]  .
    \nonumber \\  & & 
    \label{eq:U0Minkowski}
\end{eqnarray}
The background electromagnetic potential has clearly broken the maximal symmetry of the underlying Minkowski space-time, and $U_{0}^{\rm {M}}$ depends on the direction in which the space-time points $x$, $x'$ are separated. 

\subsection{$U_{1}$}
\label{sec:U1}

The equation (\ref{eq:Uneqn}) satisfied by $U_{1}$ is
\begin{equation}
  \left[ D_{\mu }D^{\mu }- m^{2}-\xi R\right] U_{0}
  - 
  \left[ \sigma ^{;\mu }D_{\mu }
 +1 -\Delta ^{-\frac{1}{2}}\Delta ^{\frac{1}{2}}_{;\mu }\sigma ^{;\mu } \right] U_{1} =0 .
 \label{eq:U1eqn}
\end{equation}
Although this equation is more complicated than that for $U_{0}$ (\ref{eq:U0eqn}), we only require $U_{1}$ to order ${\mathcal {O}}(\sigma ^{\frac{3}{2}})$.
In four space-time dimensions, the Hadamard parametrix involves terms proportion to $\ln \sigma $, which are multiplied by a biscalar $V(x,x')=V_{0}(x,x')+V_{1}(x,x')\sigma + \ldots $.
The equation (\ref{eq:U1eqn}) is similar in form (but with different numerical coefficients) to that satisfied by the Hadamard coefficient $V_{0}$ in four space-time dimensions \cite{Balakumar:2019djw}.
We therefore expect that our solution for $U_{1}$ here will feature some terms similar to those arising in $V_{0}$, which was calculated to order ${\mathcal {O}}(\sigma )$ in \cite{Balakumar:2019djw}.  

Before we can solve (\ref{eq:U1eqn}), we require a compact expression for $D^{\alpha }D_{\alpha }U_{0}$.
This is a rather lengthy calculation, but can be simplified by first writing $U_{0}$ as \cite{Balakumar:2019djw}
\begin{equation} 
U_{0}= \Delta ^{\frac{1}{2}} + {\widetilde {U}}_{0},
\end{equation}
where ${\widetilde {U}}_{0}$ vanishes when $q=0$.
Using the linearity of the covariant derivative, we then have
\begin{equation}
    D_{\alpha } U_{0} = \Delta ^{\frac{1}{2}}_{;\alpha } - {\rm {i}}q A_{\alpha }\Delta ^{\frac{1}{2}} + D_{\alpha } {\widetilde {U}}_{0},
    \label{eq:DaU01}
\end{equation}
where the expansion of $\Delta ^{\frac{1}{2}}_{;\alpha }$ to the required order can be found in (\ref{eq:DeltaDexpansion}).
Since we have found $U_{0}$ to order ${\mathcal {O}}(\sigma ^{\frac{5}{2}})$, we can compute $D_{\alpha }U_{0}$ to order ${\mathcal {O}}(\sigma ^{2})$.
The quantity $D_{\alpha }{\widetilde{U}}_{0}$ has the form  (\ref{eq:Dalphasimp}) but with the coefficients $U_{01\mu }$,\ldots replaced by ${\widetilde {U}}_{01\mu }$, \ldots .
This gives
\begin{eqnarray}
    D_{\alpha }{\widetilde {U}}_{0} & = &
    {\mathfrak {U}}_{00\alpha } +{\mathfrak {U}}_{01\alpha \mu }\sigma ^{;\mu } + {\mathfrak {U}}_{02\alpha \mu \nu }\sigma ^{;\mu \nu }
    +{\mathfrak {U}}_{03\alpha \mu \nu \lambda }\sigma ^{;\mu }\sigma ^{;\nu }\sigma ^{;\lambda }
    + {\mathfrak {U}}_{04\alpha \mu \nu \lambda \tau } \sigma ^{;\mu }\sigma ^{;\nu }\sigma ^{;\lambda }\sigma ^{;\tau }
    \nonumber \\ & & 
    + \ldots ,
    \label{eq:U0TildeDseries}
\end{eqnarray}
where the first few terms in the expansion are
\begin{subequations}
\label{eq:U0TildeD}
\begin{eqnarray}
{\mathfrak {U}}_{00\alpha } & = & 
{\rm {i}}q A_{\alpha } ,
\\
{\mathfrak {U}}_{01\alpha \mu } & = &
\frac{{\rm {i}}q}{2} F_{\alpha \mu} ,
\\
{\mathfrak {U}}_{02\alpha \mu \nu } & = & 
\frac{{\rm {i}}q}{6}\nabla _{(\mu }F_{\nu )\alpha }
+ \frac{{\rm {i}}q}{4} A_{(\alpha }R_{\mu \nu )}
+\frac{q^{2}}{2} A_{(\mu }F_{\nu )\alpha } ,
\\ 
{\mathfrak {U}}_{03\alpha \mu \nu \lambda } & = & 
-\frac{{\rm {i}}q}{24}\nabla_{(\mu } \nabla _{\nu }F_{\lambda ) \alpha }
-\frac{q^{2}}{6}A_{(\mu }\nabla _{\nu }F_{\lambda ) \alpha } 
+ \frac{{\rm {i}}q}{24} F_{\alpha (\mu }R_{\nu \lambda )}
+ \frac{q^{2}}{4}F_{\alpha (\mu }D_{\nu }A_{\lambda )}
\nonumber \\ & & 
- \frac{{\rm {i}}q}{24}R^{\rho }{}_{(\mu |\alpha |\nu }F_{\lambda ) \rho }
- \frac{{\rm {i}}q}{12}R_{\alpha (\mu }D_{\nu }A_{\lambda )}
- \frac{{\rm {i}}q}{12} R_{\alpha (\mu ;\nu }A_{\lambda )}
+ \frac{{\rm {i}}q}{24}A_{(\mu }R_{\nu \lambda );\alpha }
\nonumber \\  & & 
- \frac{{\rm {i}}q}{24}R_{(\mu \nu ;\lambda )} A_{\alpha }.
\end{eqnarray}
\end{subequations}
The expression for ${\mathfrak {U}}_{04\alpha \mu \nu \lambda \tau }$ contains many terms, so we do not reproduce it here in its entirety.
Most of these terms involve combinations of curvature tensors with either the electromagnetic potential or gauge field strength.  
In Minkowski space-time, these terms all vanish identically and 
${\mathfrak {U}}_{04\alpha \mu \nu \lambda \tau }$ reduces to
\begin{eqnarray}
  {\mathfrak {U}}_{04\alpha \mu \nu \lambda \tau }^{\rm {M}} 
 &  = & \frac{{\rm {i}}q}{120} \nabla _{(\mu }\nabla _{\nu }\nabla _{\lambda } F_{\tau )\alpha }
  + \frac{q^{2}}{24} A_{(\mu }\nabla _{\nu }\nabla _{\lambda }F_{\tau ) \alpha }
  - \frac{q^{2}}{12} F_{\alpha (\mu }D_{\nu }D_{\lambda }A_{\tau )}
  \nonumber \\ & & 
  + \frac{q^{2}}{12} \left( D_{(\mu } A_{\nu }\right) \nabla _{\lambda }F_{\tau ) \alpha }.
\end{eqnarray}
The expansion coefficients ${\mathfrak {U}}_{01\alpha \mu }$, ${\mathfrak {U}}_{02\alpha \mu \nu }$, etc.~are symmetric in the indices $\mu $, $\nu $, etc.~but the index $\alpha $ is not included in the symmetrization.
The expressions (\ref{eq:U0TildeD}) involve curvature tensors, the electromagnetic potential $A_{\mu }$ and also the field strength $F_{\mu \nu }$. 
The most compact expressions we have been able to find involve gauge covariant derivatives $D_{\mu }$ acting on the electromagnetic potential $A_{\mu }$ and space-time covariant derivatives $\nabla _{\mu }$ acting on the electromagnetic field strength $F_{\mu \nu }$.
The expansion coefficients (\ref{eq:U0TildeD}) simplify greatly in Minkowski space-time, when all curvature tensors vanish, and we are left with
\begin{eqnarray}
D_{\alpha }U_{0}^{\rm {M}} & = & 
\frac{{\rm {i}}q}{2} F_{\alpha \mu} \sigma ^{;\mu }
+ \left[ \frac{{\rm {i}}q}{6}\nabla _{(\mu }F_{\nu )\alpha }
+\frac{q^{2}}{2} A_{(\mu }F_{\nu )\alpha }  \right] \sigma ^{;\mu }\sigma ^{;\nu }
\nonumber \\ & & 
+ \left[
-\frac{{\rm {i}}q}{24}\nabla_{(\mu } \nabla _{\nu }F_{\lambda ) \alpha }
-\frac{q^{2}}{6}A_{(\mu }\nabla _{\nu }F_{\lambda ) \alpha } 
+ \frac{q^{2}}{4}F_{\alpha (\mu }D_{\nu }A_{\lambda )}
\right] \sigma ^{;\mu }\sigma ^{;\nu }\sigma ^{;\lambda }
\nonumber \\ & & 
+ \left[ 
\frac{{\rm {i}}q}{120} \nabla _{(\mu }\nabla _{\nu }\nabla _{\lambda } F_{\tau )\alpha }
  + \frac{q^{2}}{24} A_{(\mu }\nabla _{\nu }\nabla _{\lambda }F_{\tau ) \alpha }
  - \frac{q^{2}}{12} F_{\alpha (\mu }D_{\nu }D_{\lambda }A_{\tau )}
  \right. \nonumber \\ & & \left.
  + \frac{q^{2}}{12} \left[ D_{(\mu } A_{\nu }\right] \nabla _{\lambda }F_{\tau ) \alpha }
\right] \sigma ^{;\mu }\sigma ^{;\nu }\sigma ^{;\lambda }\sigma ^{;\tau } 
+ \ldots 
\label{eq:DaU0Minkowski}
\end{eqnarray}
Note that the lowest-order terms coming from $-{\rm {i}}q\Delta ^{\frac{1}{2}}$ and $D_{\alpha }{\widetilde {U}}_{0}$ have cancelled.

To find $D^{\alpha }D_{\alpha }U_{0}$, we take the derivative of (\ref{eq:DaU01}), yielding
\begin{eqnarray}
D^{\alpha }D_{\alpha }U_{0} & = & 
D^{\alpha }D_{\alpha } \Delta ^{\frac{1}{2}} 
+ D^{\alpha }D_{\alpha }{\widetilde{U}}_{0}
\nonumber \\  & = & 
\Box \Delta ^{\frac{1}{2}} -2{\rm {i}}q A^{\alpha }\Delta ^{\frac{1}{2}}_{;\alpha }- {\rm {i}}q \Delta ^{\frac{1}{2}} D^{\alpha }A_{\alpha } + D^{\alpha }D_{\alpha }{\widetilde {U}}_{0}.
\label{eq:DaDaU0}
\end{eqnarray}
We require this to order ${\mathcal {O}}(\sigma ^{\frac{3}{2}})$.
The first three terms in (\ref{eq:DaDaU0}) can be easily found using the expansions (\ref{eq:Deltahalf}, \ref{eq:boxDelta}, \ref{eq:DeltaDexpansion}), giving
\begin{equation}
\label{eq:boxdeltapart}
\Box \Delta ^{\frac{1}{2}} -2{\rm {i}}q A^{\alpha }\Delta ^{\frac{1}{2}}_{;\alpha }- {\rm {i}}q \Delta ^{\frac{1}{2}} D^{\alpha }A_{\alpha } = {\mathcal {D}}_{0}
 +{\mathcal {D}}_{1\mu }\sigma ^{;\mu }
 +{\mathcal {D}}_{2\mu \nu }\sigma ^{;\mu }\sigma ^{;\nu }
 + {\mathcal {D}}_{3\mu \nu \lambda }\sigma ^{;\mu }\sigma ^{;\nu }\sigma ^{;\lambda } 
 +\ldots 
\end{equation}
where
\begin{subequations}
\begin{eqnarray}
 {\mathcal {D}}_{0} &  = & 
 \frac{1}{6}R -{\rm {i}}q D^{\alpha }A_{\alpha } ,
 \\
 {\mathcal {D}}_{1\mu } & = & 
 -\frac{{\rm {i}}q}{3} A^{\alpha}  R_{\alpha \mu }, 
 \\
 {\mathcal {D}}_{2\mu \nu } & = & 
 \frac{1}{40} \Box R_{\mu\nu }
- \frac{1}{120} R_{;\mu\nu }
+ \frac{1}{72} R R_{\mu\nu }
- \frac{1}{30} R^{\rho }{}_{\mu}R_{\rho \nu }
+ \frac{1}{60} R^{\rho \tau }R_{\rho \mu\tau \nu }
 \nonumber \\ & & 
+ \frac{1}{60} R^{\rho \kappa \tau }{}_{\mu}R_{\rho \kappa \tau \nu }
-\frac{{\rm {i}}q}{12} A^{\alpha } R_{\mu \nu ;\alpha  } 
+ \frac{{\rm {i}}q}{6} A^{\alpha } R_{\alpha (\mu ;\nu )} 
- \frac{{\rm {i}}q}{12}R_{\mu \nu }D^{\alpha }A_{\alpha } , 
\nonumber \\
 \\
  {\mathcal {D}}_{3\mu \nu \lambda } & = & 
  \frac{1}{360} R_{;(\mu\nu \lambda )}
- \frac{1}{120} \left[ \Box R_{(\mu\nu } \right] _{;\lambda )}
- \frac{1}{144} R R_{(\mu\nu ;\lambda )}
+ \frac{1}{45} R^{\rho }{}_{(\mu}R_{|\rho |\nu ;\lambda )} 
 \nonumber \\ & &  
- \frac{1}{180} R^{\rho }{}_{\tau ; (\mu}R^{\tau }{}_{\nu |\rho |\lambda )}
- \frac{1}{180} R^{\rho }{}_{\tau } R^{\tau }{}_{(\mu |\rho |\nu ; \lambda ) }
- \frac{1}{90} R^{\rho \kappa \tau }{}_{(\mu}R_{|\rho \kappa \tau |\nu ;\lambda )}
\nonumber \\ & & 
-\frac{{\rm {i}}q}{20} A^{\alpha } R_{\alpha  (\mu ; \nu \lambda )}
+\frac{{\rm {i}}q}{30} A^{\alpha } R_{(\mu \nu ;|\alpha |\lambda ) }
-\frac{{\rm {i}}q}{36}A^{\alpha } R_{\alpha (\mu }R_{\nu \lambda )}
 \nonumber \\ & &  
- \frac{{\rm {i}}q}{45} A^{\alpha } R^{\rho }{}_{\alpha  \tau (\mu  }R^{\tau }{}_{\nu |\rho | \lambda )}
+ \frac{{\rm {i}}q}{180} A^{\alpha } R_{\rho (\mu}R^{\rho }{}_{\nu \lambda )\alpha }
+ \frac{{\rm {i}}q}{24} R_{(\mu \nu ;\lambda )} D^{\alpha }A_{\alpha } .
 \nonumber \\ & & 
\end{eqnarray}
\end{subequations}
In Minkowski spacetime, the quantities $\Delta ^{\frac{1}{2}}_{{\rm {M}};\mu }$ (\ref{eq:DeltaDexpansion}) and  $\Box \Delta _{\rm {M}}^{\frac{1}{2}}$ (\ref{eq:boxDelta}) vanish identically, we have $\Delta _{\rm {M}}^{\frac{1}{2}}=1$ and (\ref{eq:boxdeltapart}) simplifies to
\begin{equation}
D^{\alpha }D_{\alpha } \Delta ^{\frac{1}{2}}_{\rm {M}}= \Box \Delta _{\rm {M}} ^{\frac{1}{2}} -2{\rm {i}}q A^{\alpha }\Delta ^{\frac{1}{2}}_{{\rm {M}};\alpha }- {\rm {i}}q \Delta ^{\frac{1}{2}}_{\rm {M}} D^{\alpha }A_{\alpha }=   -{\rm {i}}q D^{\alpha }A_{\alpha } .
 \label{eq:boxdeltapartM}
\end{equation}
This is an exact expression in Minkowski space-time and is nonzero because we are considering gauge covariant derivatives.

This leaves $D^{\alpha }D_{\alpha }{\widetilde {U}}_{0}$ to be computed. 
Taking the covariant derivative of $D_{\alpha }{\widetilde{U}}_{0}$ (\ref{eq:U0TildeDseries}) and using the expansion (\ref{eq:sigmaexpansion}), we find
\begin{equation}
 D_{\alpha  }D^{\alpha  }{\widetilde {U}}_{0} = {\mathcal {U}}_{00}
 +{\mathcal {U}}_{01\mu }\sigma ^{;\mu }
 +{\mathcal {U}}_{02\mu \nu }\sigma ^{;\mu }\sigma ^{;\nu }
 + {\mathcal {U}}_{03\mu \nu \lambda }\sigma ^{;\mu }\sigma ^{;\nu }\sigma ^{;\lambda } 
 +\ldots 
 \label{eq:UtildeDDseries}
\end{equation}
where the coefficients ${\mathcal {U}}_{0}$, ${\mathcal {U}}_{01\mu }$, \ldots , are given by 
\begin{subequations}
\label{eq:Ucoeffs}
\begin{eqnarray}
{\mathcal {U}}_{00} &  = & 
g^{\alpha \beta} {\mathfrak {U}}_{01\alpha \beta} +D^{\alpha }{\mathfrak {U}}_{00\alpha } ,
\\ 
{\mathcal {U}}_{01\mu } & = & 
2g^{\alpha \beta }{\mathfrak {U}}_{02\alpha \beta \mu }
+
D^{\alpha }{\mathfrak {U}}_{01\alpha \mu } ,
\\ 
{\mathcal {U}}_{02\mu \nu } & =  &
 3g^{\alpha \beta }{\mathfrak {U}}_{03\alpha \beta \mu \nu }
+ D^{\alpha }{\mathfrak {U}}_{02\alpha \mu \nu } 
-\frac{1}{3} R^{\alpha }{}_{\mu }{}^{\beta }{}_{\nu }{\mathfrak {U}}_{01\alpha \beta },
 \\
 {\mathcal {U}}_{03\mu \nu \lambda } & = &
 4 g^{\alpha \beta }{\mathfrak {U}}_{04\alpha \beta \mu \nu \lambda }
 + D^{\alpha }{\mathfrak {U}}_{03\alpha \mu \nu \lambda }
 -\frac{2}{3}R^{\alpha }{}_{\mu }{}^{\beta }{}_{\nu }{\mathfrak {U}}_{02\alpha \beta \lambda }
 + \frac{1}{12}R^{\alpha }{}_{\mu }{}^{\beta }{}_{\nu ;\lambda }{\mathfrak {U}}_{01\alpha \beta }.
 \nonumber \\ & & 
\end{eqnarray}
\end{subequations}
Substituting in from (\ref{eq:U0TildeD}), and simplifying, we find that the first three terms in the expansion (\ref{eq:UtildeDDseries}) are
\begin{subequations}
\label{eq:U0TildeDD}
\begin{eqnarray}
{\mathcal {U}}_{00} &  = & 
{\rm {i}}qD^{\mu }A_{\mu } ,
\label{eq:U0simp}
\\
{\mathcal {U}}_{01\mu } & = & 
\frac{{\rm {i}}q}{3} \left[ A^{\alpha }R_{\alpha \mu } + \frac{1}{2}RA_{\mu }
+ \nabla ^{\alpha }F_{\alpha \mu }
\right]  ,
\label{eq:U1simp}
\\ 
{\mathcal {U}}_{02\mu \nu } & =  &
\frac{{\rm {i}}q}{12} R_{\mu \nu }D_{\alpha }A^{\alpha }
- \frac{{\rm {i}}q}{12} R D_{(\mu }A_{\nu )}
+ \frac{{\rm {i}}q}{12} A^{\alpha }R_{\mu \nu ;\alpha}
- \frac{{\rm {i}}q}{6}A^{\alpha }R_{\alpha (\mu ;\nu )}
\nonumber \\ & & 
+ \frac{q^{2}}{4}F^{\alpha }{}_{\mu }F_{\nu \alpha }
+ \frac{{\rm {i}}q}{12}\nabla ^{\alpha }\nabla _{(\mu }F_{\nu )\alpha }
+ \frac{q^{2}}{3}A_{(\mu }\nabla ^{\alpha }F_{\nu )\alpha }
- \frac{{\rm {i}}q}{12}R^{\alpha }{}_{(\mu }F_{\nu ) \alpha } .
\nonumber \\ & & 
\label{eq:U2simp}
\end{eqnarray}
\end{subequations}
Again we do not give explicitly the rather lengthy expression for ${\mathcal {U}}_{03\mu \nu \lambda } $,
except in the particular case of Minkowkski space-time, when all the curvature tensors vanish identically, and we have the simplified result
\begin{eqnarray}
{\mathcal {U}}_{03\mu \nu \lambda }^{\rm {M}} & = & 
-\frac{{\rm {i}}q}{60} \nabla ^{\alpha }\nabla _{(\mu }\nabla _{\nu }F_{\lambda ) \alpha }
-\frac{q^{2}}{12} A_{(\mu }\nabla ^{\alpha }\nabla _{\nu }F_{\lambda ) \alpha }
- \frac{q^{2}}{6}\left[ D_{(\mu }A_{\nu } \right] \nabla ^{\alpha }F_{\lambda ) \alpha }
\nonumber \\ & & 
-\frac{q^{2}}{6}F^{\alpha }{}_{(\mu }\nabla _{\nu }F_{\lambda ) \alpha }
- \frac{{\rm {i}}q^{3}}{4} F_{\alpha (\mu }F^{\alpha }{}_{\nu }A_{\lambda )}
.
\label{eq:U3simp}
\end{eqnarray}
In Minkowski space-time, the earlier terms in the expansion (\ref{eq:U0TildeDD}) also simplify and we find
\begin{eqnarray}
D^{\alpha }D_{\alpha }U_{0}^{\rm {M}} & = & 
 \frac{{\rm {i}}q}{3}\left[ \nabla ^{\alpha }F_{\alpha \mu } \right] \sigma ^{;\mu }
\nonumber \\ & & 
+ \left[ 
\frac{q^{2}}{4}F^{\alpha }{}_{\mu }F_{\nu \alpha }
+ \frac{{\rm {i}}q}{12}\nabla ^{\alpha }\nabla _{(\mu }F_{\nu )\alpha }
+ \frac{q^{2}}{3}A_{(\mu }\nabla ^{\alpha }F_{\nu )\alpha }
\right] \sigma ^{;\mu }\sigma ^{;\nu }
\nonumber \\ & & 
+ \left[ 
-\frac{{\rm {i}}q}{60} \nabla ^{\alpha }\nabla _{(\mu }\nabla _{\nu }F_{\lambda ) \alpha }
-\frac{q^{2}}{12} A_{(\mu }\nabla ^{\alpha }\nabla _{\nu }F_{\lambda ) \alpha }
- \frac{q^{2}}{6}\left[ D_{(\mu }A_{\nu } \right] \nabla ^{\alpha }F_{\lambda ) \alpha }
\right. \nonumber \\ & & \left.  \qquad
-\frac{q^{2}}{6}F^{\alpha }{}_{(\mu }\nabla _{\nu }F_{\lambda ) \alpha }
- \frac{{\rm {i}}q^{3}}{4} F_{\alpha (\mu }F^{\alpha }{}_{\nu }A_{\lambda )}
\right] \sigma ^{;\mu }\sigma ^{;\nu } \sigma ^{;\lambda }
\ldots .
\end{eqnarray}
In this case the zeroth order terms arising from (\ref{eq:boxdeltapartM}, \ref{eq:U0simp}) have cancelled, leaving $D^{\alpha }D_{\alpha }U_{0}^{\rm {M}}$ to be ${\mathcal {O}}(\sigma ^{\frac{1}{2}})$.

We now have all the ingredients needed to solve (\ref{eq:U1eqn}).
The quantity $D_{\alpha }U_{1}$ has a similar form to (\ref{eq:Dalphasimp}), but with the $U_{0}$ coefficients replaced by $U_{1}$ coefficients.
The method is then similar to that used to find the expansion of $U_{0}$.
Setting the coefficients of the series expansion resulting from (\ref{eq:U1eqn}) to zero, we obtain the equations
\begin{subequations}
\label{eq:U1eqns}
\begin{eqnarray}
0 & = & 
U_{10}-{\mathcal {D}}_{0} -{\mathcal {U}}_{00} + \left( m^{2}+\xi R \right) U_{00},
\label{eq:U10eqn}
\\
0 & = & 
 2U_{11\alpha }+ D_{\alpha }U_{10} - {\mathcal {D}}_{1\alpha  }
 -{\mathcal {U}}_{01\alpha }
 + \left( m^{2}+\xi R \right) U_{01\alpha  },
\label{eq:U11eqn}
\\
0 & = & 
3U_{12\alpha \mu } + D_{(\alpha  }U_{11\mu )}  - \frac{1}{6}R_{\mu \alpha}U_{10} -{\mathcal {D}}_{2\alpha  \mu  }
- {\mathcal {U}}_{02\alpha \mu }
+ \left( m^{2}+\xi R \right) U_{02\alpha \mu },
\nonumber \\ & & 
\label{eq:U12eqn}
\\
0 & = & 
4U_{13\alpha \mu \nu }+ D_{(\alpha }U_{12\mu \nu )} 
  - \frac{1}{6}R_{(\mu \alpha}U_{11 \nu )}
 + \frac{1}{24} R_{(\mu \nu ; \alpha )}U_{10} - {\mathcal {D}}_{3\alpha \mu \nu } 
 - {\mathcal {U}}_{03\alpha \mu \nu }
  \nonumber \\ & & 
 + \left( m^{2}+\xi R \right) U_{03\alpha \mu \nu }.
\label{eq:U13eqn}
\end{eqnarray}
\end{subequations}
We now proceed as we did for $U_{0}$ and solve the equations (\ref{eq:U1eqns}) iteratively.
The first equation (\ref{eq:U10eqn}) is straightforward to solve:
\begin{subequations}
\label{eq:U1expansionsimp}
\begin{equation}
U_{10}  =  
{\mathcal {D}}_{0} +{\mathcal {U}}_{00}-  \left( m^{2}+\xi R \right) U_{00}= 
- \left[ m^{2}+  \left( \xi - \frac{1}{6} \right) R \right]  . 
\end{equation}
Therefore $U_{10}$ does not receive any contributions from the electromagnetic field, but does depend on the scalar field mass $m$ and the coupling $\xi $ of the scalar field to the curvature scalar $R$. 
This behaviour is shared with the zeroth order term in the covariant series expansion of $V_{0}$ in four space-time dimensions \cite{Balakumar:2019djw}. 

Next, we have 
\begin{eqnarray}
U_{11\mu } & = &  
-\frac{1}{2} \left[ D_{\alpha }U_{10} - {\mathcal {D}}_{01\alpha  }
 -{\mathcal {U}}_{01\alpha }
 + \left( m^{2}+\xi R \right) U_{01\alpha  } \right] 
 \nonumber \\ & = &
 \frac{1}{2} \left( \xi - \frac{1}{6} \right) R_{;\mu }
 -{\rm {i}}q \left[ m^{2} + \left( \xi - \frac{1}{6} \right) R \right] A_{\mu }
 + \frac{{\rm {i}}q}{6}\nabla ^{\alpha }F_{\alpha \mu }.
\end{eqnarray}
Corrections due to the electromagnetic potential have arisen at this order.
Continuing to (\ref{eq:U12eqn}), after some algebra we find
\begin{eqnarray}
U_{12\mu \nu }& = & 
-\frac{1}{6}m^{2}R_{\mu \nu }
- \frac{1}{3} \left( \xi - \frac{3}{20} \right) R_{;\mu \nu }
+ \frac{1}{60} \Box R_{\mu \nu }
- \frac{1}{6} \left( \xi - \frac{1}{6} \right) R R_{\mu \nu }
\nonumber \\ & & 
-\frac{1}{45} R^{\alpha }{}_{\mu }R_{\alpha \nu }
+ \frac{1}{90} R^{\alpha \beta }R_{\alpha \mu \beta \nu }
+ \frac{1}{90} R^{\alpha \beta \gamma }{}_{\mu }R_{\alpha \beta \gamma \nu }
\nonumber \\  & & 
+\frac{{\rm {i}}q}{2} \left[ m^{2} + \left( \xi - \frac{1}{6} \right) R \right] D_{(\mu }A_{\nu  )}
+ \frac{{\rm {i}}q}{2} \left( \xi - \frac{1}{6} \right) A_{(\mu }R_{;\nu )}
\nonumber \\  & & 
+\frac{q^{2}}{12} F^{\alpha }{}_{\mu }F_{\nu \alpha }
+ \frac{q^{2}}{6} A_{(\mu }\nabla ^{\alpha }F_{\nu ) \alpha }
+ \frac{{\rm {i}}q}{12} \nabla _{(\mu }\nabla ^{\alpha }F_{\nu ) \alpha }.
\end{eqnarray}
\end{subequations}
This coefficient is considerably more complicated than the previous two, and involves coupling between the electromagnetic potential and curvature tensors, as well as higher-order derivatives of the electromagnetic field.

As anticipated from our calculations of 
$ {\mathfrak {U}}_{04 \alpha \mu \nu \lambda }$
and $ {\mathcal {U}}_{03\alpha \mu \nu }$, we find that $U_{13\mu \nu \lambda }$ involves many complicated terms coupling electromagnetic field quantities and curvature tensors.
We therefore restrict ourselves to giving the form of $U_{13\mu \nu \lambda }$ in Minkowski space-time, which is
\begin{eqnarray}
U_{13\mu \nu \lambda }^{\rm {M}} & = &
-\frac{{\rm {i}}qm^{2}}{6} D_{(\mu }D_{\nu } A_{\lambda ) }
- \frac{{\rm {i}}q}{40} \nabla ^{\alpha }\nabla _{(\mu }\nabla _{\nu }F_{\lambda ) \alpha }
- \frac{q^{2}}{12}\left[ D_{(\mu }A_{\nu }\right] \nabla ^{\alpha }F_{\lambda ) \alpha }
\nonumber \\ & & 
+ \frac{{\rm {i}}q^{3}}{12}A_{(\mu }F^{\alpha }{}_{\nu }F_{\lambda ) \alpha }
- \frac{q^{2}}{12} F^{\alpha }{}_{(\mu }\nabla _{\nu }F_{\lambda ) \alpha }
-\frac{q^{2}}{12} A_{(\mu }\nabla ^{\alpha }\nabla _{\nu }F_{\lambda ) \alpha }.
\end{eqnarray}
The lower-order terms in the expansion (\ref{eq:U1expansionsimp}) also simplify in Minkowski space-time, to yield 
\begin{eqnarray}
U_{1}^{\rm {M}}  & = & 
-m^{2} 
+ \left[  
 -{\rm {i}}q m^{2} A_{\mu }
 + \frac{{\rm {i}}q}{6}\nabla ^{\alpha }F_{\alpha \mu }
 \right] \sigma ^{;\mu }
 \nonumber \\ & & 
 + \left[
\frac{{\rm {i}}q}{2} m^{2} D_{(\mu }A_{\nu  )}
+\frac{q^{2}}{12} F^{\alpha }{}_{\mu }F_{\nu \alpha }
+ \frac{q^{2}}{6} A_{(\mu }\nabla ^{\alpha }F_{\nu ) \alpha }
+ \frac{{\rm {i}}q}{12} \nabla _{(\mu }\nabla ^{\alpha }F_{\nu ) \alpha } 
 \right] \sigma ^{;\mu }\sigma ^{;\nu }
 \nonumber \\ & & 
+ \left[
 -\frac{{\rm {i}}qm^{2}}{6} D_{(\mu }D_{\nu } A_{\lambda ) }
- \frac{{\rm {i}}q}{40} \nabla ^{\alpha }\nabla _{(\mu }\nabla _{\nu }F_{\lambda ) \alpha }
- \frac{q^{2}}{12}\left[ D_{(\mu }A_{\nu }\right] \nabla ^{\alpha }F_{\lambda ) \alpha }
\right. \nonumber \\ & & \left. \quad 
+ \frac{{\rm {i}}q^{3}}{12}A_{(\mu }F^{\alpha }{}_{\nu }F_{\lambda ) \alpha }
- \frac{q^{2}}{12} F^{\alpha }{}_{(\mu }\nabla _{\nu }F_{\lambda ) \alpha }
-\frac{q^{2}}{12} A_{(\mu }\nabla ^{\alpha }\nabla _{\nu }F_{\lambda ) \alpha }
 \right] \sigma ^{;\mu }\sigma ^{;\nu }\sigma ^{;\lambda }
 +\ldots 
 \nonumber \\ & & 
 \label{eq:U1Minkowski}
\end{eqnarray}
Even in Minkowski space-time, the expression (\ref{eq:U1Minkowski}) for $U_{1}^{\rm {M}}$ is considerably more complicated than the corresponding expression (\ref{eq:U0Minkowski}) for $U_{0}^{\rm {M}}$.
It depends on the scalar field mass as well as the electromagnetic potential and is nonzero if the scalar field is neutral.
If we consider a massless charged scalar field, the zeroth order term in $U_{1}^{\rm {M}}$ vanishes and $U_{1}^{\rm {M}}$ is order ${\mathcal {O}}(\sigma ^{\frac{1}{2}})$.

\subsection{$U_{2}$}
\label{sec:U2}

$U_{2}$ satisfies the equation 
\begin{equation}
  \left[ D_{\mu }D^{\mu }- m^{2}-\xi R\right] U_{1}
  + 
  \left[ \sigma ^{;\mu }D_{\mu }
 +2 -\Delta ^{-\frac{1}{2}}\Delta ^{\frac{1}{2}}_{;\mu }\sigma ^{;\mu } \right] U_{2} =0 .
 \label{eq:U2eqn}
 \end{equation}
 This is similar in structure to the equation (\ref{eq:U1eqn}) for $U_{1}$, but we only need to find $U_{2}$ to order ${\mathcal {O}}(\sigma ^{\frac{1}{2}})$.
 Equation (\ref{eq:U2eqn}) is also similar in form (but with different numerical coefficients) to the equation governing the coefficient $V_{1}$ in four space-time dimensions \cite{Balakumar:2019djw}, although there we solved the corresponding equation only to leading order.
 
Our method follows that for finding $U_{1}$ in section \ref{sec:U1}, with our first aim to find an expression for $D^{\alpha }D_{\alpha }U_{1}$ in as compact a form as possible.
We begin by writing $D_{\alpha }U_{1}$ as
\begin{equation}
    D_{\alpha }U_{1} = {\mathfrak {U}}_{10\alpha }+ {\mathfrak {U}}_{11\alpha \mu }\sigma ^{;\mu } + {\mathfrak {U}}_{12\alpha \mu \nu } \sigma ^{;\mu }\sigma ^{;\nu } + \ldots 
    \label{eq:DaU1}
\end{equation}
The coefficients in this expansion are found using an analogous expression to  (\ref{eq:Dalphasimp}), with the $U_{0}$ coefficients replaced by $U_{1}$ coefficients.
The first two terms in (\ref{eq:DaU1}) are
\begin{subequations}
\label{eq:DaU1expansion}
\begin{eqnarray}
{\mathfrak {U}}_{10\alpha } & = & 
-\frac{1}{2} \left( \xi - \frac{1}{6} \right) R_{;\alpha }+ \frac{{\rm {i}}q}{6} \nabla ^{\rho }F_{\rho \alpha },
\\
{\mathfrak {U}}_{11\alpha \mu } & = & 
-\frac{m^{2}}{6} R_{\alpha \mu }
+ \frac{1}{6} \left( \xi - \frac{1}{5} \right) R_{;\alpha \mu }
+ \frac{1}{60} \Box R_{\alpha \mu }
\nonumber \\  & & 
- \frac{1}{6} \left( \xi - \frac{1}{6} \right) R R_{\alpha \mu }
- \frac{1}{45} R^{\rho }{}_{\alpha }R_{\rho \mu }
+ \frac{1}{90} R^{\rho \tau }R_{\rho \alpha \tau \mu }
+ \frac{1}{90} R^{\rho \tau \kappa }{}_{\alpha }R_{\rho \tau \kappa \mu }
\nonumber \\ & & 
-\frac{{\rm {i}}q}{2} \left(  \xi - \frac{1}{6} \right) A_{\mu }R_{;\alpha }
+ \frac{q^{2}}{6}F^{\rho }{}_{\alpha }F_{\mu \rho }
- \frac{{\rm {i}}q}{2} \left[ m^{2} + \left( \xi - \frac{1}{6} \right) R \right] F_{\alpha \mu }
\nonumber \\ & & 
+ \frac{q^{2}}{6}A_{\mu } \nabla ^{\rho }F_{\alpha \rho } 
+ \frac{{\rm {i}}q}{12} \nabla _{\mu }\nabla ^{\rho }F_{\alpha \rho }
- \frac{{\rm {i}}q}{12} \nabla _{\alpha }\nabla ^{\rho }F_{\mu \rho }.
\end{eqnarray}
\end{subequations}
As anticipated, these expressions involve combinations of curvature tensors, the electromagnetic potential, the gauge field strength and their derivatives. 

Since ${\mathfrak {U}}_{12\alpha \mu \nu }$ involves $U_{13\mu \nu \lambda }$, we give its (lengthy) form only in Minkowski space-time:
\begin{eqnarray}
{\mathfrak {U}}_{12\alpha \mu \nu }^{\rm {M}} & = & 
-\frac{{\rm {i}}qm^{2}}{6} \nabla _{(\mu }F_{\nu ) \alpha }
+ \frac{q^{2}m^{2}}{2} F_{\alpha (\mu }A_{\nu )}
- \frac{{\rm {i}}q}{40} \nabla ^{\rho }\nabla _{(\mu }\nabla _{\nu )} F_{\alpha \rho} 
+ \frac{{\rm{i}}q}{30} \nabla ^{\rho }\nabla _{\alpha } \nabla _{(\mu }F_{\nu ) \rho }
\nonumber \\ & & 
- \frac{q^{2}}{12} A_{(\mu }\nabla ^{\rho }\nabla _{\nu )} F_{\alpha \rho}
+ \frac{q^{2}}{12} A_{(\mu }\nabla _{|\alpha | }\nabla ^{\rho }F_{\mu ) \rho }
- \frac{q^{2}}{12} \left[ D_{(\mu }A_{\nu )} \right] \nabla ^{\rho} F_{\alpha \rho }
\nonumber \\ & & 
- \frac{q^{2}}{12} F^{\rho }{}_{\alpha }\nabla _{(\mu }F_{\nu ) \rho }
-\frac{q^{2}}{12} F^{\rho }{}_{(\mu }\nabla _{\nu )}F_{\alpha \rho} 
+ \frac{q^{2}}{12}F^{\rho }{}_{(\mu }\nabla _{|\alpha |} F_{\nu ) \rho }
+ \frac{q^{2}}{12} F_{\alpha (\mu }\nabla ^{\rho }F_{\nu ) \rho }
\nonumber \\ & & 
+ \frac{{\rm {i}}q^{3}}{6} A_{(\mu }F^{\rho }{}_{\nu )} F_{\alpha \rho }.
\end{eqnarray}
On Minkowski space-time, the lower-order terms (\ref{eq:DaU1expansion}) simplify and we have
\begin{eqnarray}
D_{\alpha }U_{1}^{\rm {M}} & = & 
\frac{{\rm {i}}q}{6} \nabla ^{\rho }F_{\rho \alpha }
+ \left[ 
 \frac{q^{2}}{6}F^{\rho }{}_{\alpha }F_{\mu \rho }
- \frac{{\rm {i}}qm^{2}}{2}  F_{\alpha \mu }
+ \frac{q^{2}}{6}A_{\mu } \nabla ^{\rho }F_{\alpha \rho } 
+ \frac{{\rm {i}}q}{12} \nabla _{\mu }\nabla ^{\rho }F_{\alpha \rho }
\right. \nonumber \\ & & \left. \qquad 
- \frac{{\rm {i}}q}{12} \nabla _{\alpha }\nabla ^{\rho }F_{\mu \rho }
\right] \sigma ^{;\mu }
\nonumber \\ & & 
+ \left[
-\frac{{\rm {i}}qm^{2}}{6} \nabla _{(\mu }F_{\nu ) \alpha }
+ \frac{q^{2}m^{2}}{2} F_{\alpha (\mu }A_{\nu )}
- \frac{{\rm {i}}q}{40} \nabla ^{\rho }\nabla _{(\mu }\nabla _{\nu )} F_{\alpha \rho} 
\right. \nonumber \\ & & \left. \qquad
+ \frac{{\rm{i}}q}{30} \nabla ^{\rho }\nabla _{\alpha } \nabla _{(\mu }F_{\nu ) \rho }
- \frac{q^{2}}{12} A_{(\mu }\nabla ^{\rho }\nabla _{\nu )} F_{\alpha \rho}
+ \frac{q^{2}}{12} A_{(\mu }\nabla _{|\alpha | }\nabla ^{\rho }F_{\mu ) \rho }
\right. \nonumber \\ & & \left. \qquad
- \frac{q^{2}}{12} \left[ D_{(\mu }A_{\nu )} \right] \nabla ^{\rho} F_{\alpha \rho }
- \frac{q^{2}}{12} F^{\rho }{}_{\alpha }\nabla _{(\mu }F_{\nu ) \rho }
-\frac{q^{2}}{12} F^{\rho }{}_{(\mu }\nabla _{\nu )}F_{\alpha \rho} 
\right. \nonumber \\ & & \left. \qquad
+ \frac{q^{2}}{12}F^{\rho }{}_{(\mu }\nabla _{|\alpha |} F_{\nu ) \rho }
+ \frac{q^{2}}{12} F_{\alpha (\mu }\nabla ^{\rho }F_{\nu ) \rho }
+ \frac{{\rm {i}}q^{3}}{6} A_{(\mu }F^{\rho }{}_{\nu )} F_{\alpha \rho }
\right] \sigma ^{;\mu} \sigma ^{;\nu }\sigma ^{;\lambda } 
\nonumber \\ & & 
+ \ldots 
\end{eqnarray}
Unlike the situation for $D_{\alpha }U_{0}^{\rm {M}}$ (\ref{eq:DaU0Minkowski}), in $D_{\alpha }U_{1}^{\rm {M}}$ the zeroth order term is nonzero and depends on the divergence of the gauge field strength. 

The next step is to find $D_{\alpha }D^{\alpha }U_{1}$, which we write, to the order required, as
\begin{equation}
    D_{\alpha }D^{\alpha }U_{1} = {\mathcal {U}}_{10} + {\mathcal {U}}_{11\mu }\sigma ^{;\mu }+ \ldots 
\end{equation}
The lowest order term ${\mathcal {U}}_{10}$ is straightforward to find:
\begin{eqnarray}
{\mathcal {U}}_{10} & = & g^{\alpha \beta }{\mathfrak {U}}_{11\alpha \beta} + D^{\alpha }{\mathfrak {U}}_{10\alpha }
\nonumber \\ & = & 
\frac{1}{90}R^{\alpha \beta \gamma \delta }R_{\alpha \beta \gamma \delta }
- \frac{1}{90} R^{\alpha \beta }R_{\alpha \beta }
-\frac{1}{3}\left( \xi - \frac{1}{5} \right) \Box R
- \frac{1}{6} \left[ m^{2}+ \left( \xi - \frac{1}{6}\right) R \right] R
\nonumber \\  & & 
- \frac{q^{2}}{6}F^{\alpha \beta }F_{\alpha \beta }.
\label{eq:calU10}
\end{eqnarray}
We see that there is just a single term arising from the contribution of the electromagnetic field.
As previously, we do not give the expression for the higher order term  ${\mathcal {U}}_{11\mu }$ on a general space-time background.
On Minkowski space-time, it takes the form:
\begin{eqnarray}
{\mathcal {U}}_{11\mu }^{\rm {M}} & = & 
\frac{{\rm {i}}qm^{2}}{3}\nabla ^{\alpha }F_{\mu \alpha }
-\frac{{\rm{i}}q}{20} \Box \nabla ^{\alpha }F_{\mu \alpha }
+ \frac{q^{2}}{6}F^{\alpha \beta }\nabla _{\alpha }F_{\mu \beta }
-\frac{{\rm {i}}q^{3}}{6}A_{\mu }F^{\alpha \beta }F_{\alpha \beta}
.
\end{eqnarray}
Most of the terms in ${\mathcal {U}}_{10}$ (\ref{eq:calU10}) also vanish in Minkowski space-time, giving
\begin{eqnarray}
D^{\alpha }D_{\alpha }U_{1}^{\rm {M}} & = & 
- \frac{q^{2}}{6}F^{\alpha \beta }F_{\alpha \beta }
+ \left[
\frac{{\rm {i}}qm^{2}}{3}\nabla ^{\alpha }F_{\mu \alpha }
-\frac{{\rm{i}}q}{20} \Box \nabla ^{\alpha }F_{\mu \alpha }
+ \frac{q^{2}}{6}F^{\alpha \beta }\nabla _{\alpha }F_{\mu \beta }
\right. \nonumber \\ & & \left.
-\frac{{\rm {i}}q^{3}}{6}A_{\mu }F^{\alpha \beta }F_{\alpha \beta}
\right] \sigma ^{;\mu }
+ \ldots. 
\end{eqnarray}
This is a comparatively compact expression, particularly given the complexity of $U_{1}^{M}$ (\ref{eq:U1Minkowski}).

We can now solve (\ref{eq:U2eqn}) iteratively to obtain the expansion coefficients in $U_{2}$. 
Substituting in the expansion for $U_{2}$ (\ref{eq:U2expansion}), 
we obtain the equations
\begin{subequations}
\begin{eqnarray}
\label{eq:U2eqns}
0 & = & 
2U_{20}+{\mathcal {U}}_{10} - \left( m^{2}+\xi R \right) U_{10},
\label{eq:U20eqn}
\\
0 & = & 
 3U_{21\alpha }+ D_{\alpha }U_{20} 
 +{\mathcal {U}}_{11\alpha }
 - \left( m^{2}+\xi R \right) U_{11\alpha  }. 
\label{eq:U21eqn}
\end{eqnarray}
\end{subequations}
The first can be straightforwardly solved for a general space-time background to give
\begin{eqnarray}
U_{20} & = & 
-\frac{1}{2} \left[ m^{2}+ \left( \xi - \frac{1}{6} \right) R \right] ^{2}
+ \frac{1}{6} \left( \xi - \frac{1}{5} \right) \Box R
+\frac{1}{180} R^{\alpha \beta }R_{\alpha \beta }
\nonumber \\ & & 
- \frac{1}{180} R^{\alpha \beta \gamma \delta }R_{\alpha \beta \gamma \delta }
+ \frac{q^{2}}{12}F_{\alpha \beta }F^{\alpha \beta }.
\label{eq:U20}
\end{eqnarray}
We find a single additional term due to the electromagnetic field, which is nonzero even in Minkowski space-time.
A similar term (but with a different numerical coefficient) also arises in the zeroth order contribution to $V_{1}$ in four space-time dimensions \cite{Balakumar:2019djw}.
In that situation this term plays an important role in contributing to the trace anomaly \cite{Balakumar:2019djw}.
Here we are working in five space-time dimensions, so there is no trace anomaly \cite{Decanini:2005eg}.

We present the form of $U_{12\alpha }$ valid on Minkowski space-time only:
\begin{eqnarray}
U_{21\mu }^{\rm {M}} & = & 
-\frac{{\rm{i}}qm^{4}}{2}A_{\mu }
- \frac{{\rm {i}}qm^{2}}{6}\nabla ^{\alpha }F_{\mu \alpha }
+ \frac{{\rm {i}}q}{60} \Box \nabla ^{\alpha }F_{\mu \alpha }
-\frac{q^{2}}{12} F^{\alpha \beta }\nabla _{\mu }F_{\alpha \beta }
+\frac{{\rm {i}}q^{3}}{12} A_{\mu }F^{\alpha \beta }F_{\alpha \beta }.
\nonumber \\ & & 
\end{eqnarray}
Together with the simplified expression for $U_{20}$ (\ref{eq:U20}) in Minkowski space-time, we find
\begin{eqnarray}
U_{2}^{\rm {M}} &  = & 
\frac{q^{2}}{12}F_{\alpha \beta }F^{\alpha \beta }
+ \left[ 
-\frac{{\rm{i}}qm^{4}}{2}A_{\mu }
- \frac{{\rm {i}}qm^{2}}{6}\nabla ^{\alpha }F_{\mu \alpha }
+ \frac{{\rm {i}}q}{60} \Box \nabla ^{\alpha }F_{\mu \alpha }
-\frac{q^{2}}{12} F^{\alpha \beta }\nabla _{\mu }F_{\alpha \beta }
\right. \nonumber \\ & & \left.
+\frac{{\rm {i}}q^{3}}{12} A_{\mu }F^{\alpha \beta }F_{\alpha \beta }
\right] \sigma ^{;\mu } + \ldots .
\end{eqnarray}
The zeroth order term in $U_{2}^{\rm {M}}$ is nonzero even if the scalar field is massless, in contrast to $U_{1}^{\rm {M}}$ (\ref{eq:U1Minkowski}).

\section{Conclusions}
\label{sec:conc}

In this article we have derived the Hadamard parametrix for the Feynman Green's function of a massive, charged, complex scalar field on a five-dimensional curved space-time. 
We have presented explicit expressions for covariant Taylor series expansions to sufficiently high order as required for the computation of the renormalized current on a general space-time background, and the renormalized stress-energy tensor on Minkowski space-time.
We make no assumptions about either the background metric or electromagnetic potential.
In particular, we do not assume any form of the field equations, although our expressions would simplify if, for example, it is assumed that both Einstein's and Maxwell's equations are satisfied by the background quantities.

This work generalizes previous results for the Hadamard parametrix for a neutral scalar field in four \cite{Adler:1976jx,Wald:1978pj,Castagnino:1984mk,Brown:1986tj,Bernard:1986vc,Tadaki:1987dq} and higher dimensions \cite{Decanini:2005eg}, and the covariant Taylor series expansions for a charged scalar field in two, three and four dimensions given in \cite{Balakumar:2019djw}.
These results will be useful for the computation of renormalized expectation values for a charged scalar field on a curved space-time background.
For example, the renormalized scalar field condensate for a neutral scalar field on a five-dimensional Schwarzschild-Tangherlini black hole was computed in \cite{Taylor:2016edd}, and it would be interesting to investigate the effect of both black hole and scalar field charge on this quantity.

Using the Hadamard parametrix to regularize the Feynman Green's function for a charged scalar field is not the only approach to the renormalization of expectation values of observables.
Previous work studying the Green's function for a charged scalar field on four space-time dimensions considered the DeWitt-Schwinger representation of the Feynman Green's function \cite{Boulware:1978hy,Herman:1995hm}.
For a neutral scalar field, it is known that the singular part of the DeWitt-Schwinger representation of the Feynman Green's function is the same as that given by the Hadamard parametrix \cite{Decanini:2005gt} and we anticipate that the same is true for a charged scalar field.

The work of \cite{Balakumar:2019djw,Boulware:1978hy,Herman:1995hm}, as here, is applicable to any space-time geometry and background electromagnetic potential.  
For particular space-times, other methods of renormalization are available.
For example, for a Friedman-Lema\^itre-Robertson-Walker space-time, adiabatic regularization can be used to find renormalized expectation values for the adiabatic vacuum state.
This approach has been applied to a charged scalar field in two \cite{Ferreiro:2018qzr} and four \cite{Ferreiro:2018qdi} dimensions.
If one considers a neutral scalar field, then adiabatic regularization is equivalent to DeWitt-Schwinger (and hence Hadamard) renormalization 
\cite{Birrell:1978,delRio:2014bpa}.
Since the calculations involved in showing this are extremely lengthy, even for a neutral scalar field, we leave the interesting question of the equivalence of adiabatic and Hadamard renormalization for a charged scalar field to future work.

\section*{Acknowledgments}

We thank the organizers of the Vth Amazonian Symposium on Physics for the opportunity to present our work and for a very stimulating conference. 
This work was partially supported by the H2020-MSCA-RISE-2017 Grant No.~FunFiCO-777740.
V.B.~thanks STFC for the provision of a studentship supporting this work and the Faculdade de F\'isica, 
Universidade Federal do Par\'a, Bel\'em, Par\'a, Brazil for warm and generous hospitality. 
The work of E.W.~is also supported by the Lancaster-Manchester-Sheffield Consortium for
Fundamental Physics under STFC grant ST/P000800/1.

\end{document}